\documentclass[aps,prd,a4paper,showpacs,12pt]{article}
 
\usepackage[margin=1.0in]{geometry}
\usepackage{multicol}
\usepackage{subfigure}
\usepackage{graphicx}
\usepackage{color}
\usepackage{xcolor}
\usepackage{threeparttable}
\usepackage{amsfonts}
\usepackage{times}
\usepackage{setspace}
\usepackage{balance}
\usepackage{lastpage}
\usepackage{amsthm}
\usepackage[tbtags]{amsmath}
\usepackage{nomencl}
\usepackage{pstricks}
\usepackage{pstricks-add}
\usepackage{pst-plot,pstricks-add}

\usepackage{amsmath,amsfonts,amssymb,simplewick}
\usepackage{float}

\usepackage{amsmath,amssymb,amsfonts,epsfig,graphicx,euscript}%
\usepackage{hyperref}
\hypersetup{bookmarks=true, bookmarksnumbered=true,linktoc=page, pdfstartview={FitH}, 
colorlinks=true, 
citecolor=red, 
filecolor=blue, 
linkcolor=blue, 
urlcolor=blue}
\usepackage{amsmath}
\usepackage{amsfonts}
\usepackage{graphicx}
\usepackage{caption}
\usepackage{float}
\usepackage[all]{hypcap}
\numberwithin{equation}{section}
\usepackage{cite}
\usepackage{calc}
\newlength{\spacer}
\newsavebox{\mybox}

\newcommand{\bse}{\begin{subequations}}
 \newcommand{\ese}{\end{subequations}}
\newcommand{\be}{\begin{equation}}
\newcommand{\ee}{\end{equation}}
\newcommand{\bea}{\begin{eqnarray}}
\newcommand{\eea}{\end{eqnarray}}
\newcommand{\ba}{\begin{array}}
 \newcommand{\ea}{\end{array}}

\renewcommand{\thefootnote}{\fnsymbol{footnote}}
\begin{document}
\begin{center}
	   { \large{\textbf{The generation of matter-antimatter asymmetries and hypermagnetic fields by the chiral vortical effect of transient fluctuations}}} 
	   	\vspace*{1.5cm}
\begin{center}
	{\bf S. Abbaslu\footnote{s$_{-}$abbasluo@sbu.ac.ir}$^1$, S. Rostam Zadeh\footnote{sh$_{-}$rostamzadeh@ipm.ir}$^2$, \bf M. Mehraeen\footnote{mxm1289@case.edu}$^{1,3}$ and S. S. Gousheh\footnote{ss-gousheh@sbu.ac.ir}$^1$}\\
	\vspace*{0.5cm}
	{\it{$^1$Department of Physics, Shahid Beheshti University, Tehran, Iran\\$^2$School of Particles and Accelerators, Institute for Research in Fundamental Sciences (IPM), P.O.Box 19395-5531, Tehran, Iran}}\\
	{\it{$^3$Department of Physics, Case Western Reserve University, 10900 Euclid Avenue, Cleveland, Ohio 44106, USA}}\\
	\vspace*{1cm}
\end{center}
	\end{center}
	\begin{center}
		\today
	\end{center}
	\renewcommand*{\thefootnote}{\arabic{footnote}}
\setcounter{footnote}{0}
	
\date{\today}
\textbf{Abstract:}
We study the contribution of the temperature-dependent chiral vortical effect to the generation and evolution of  hypermagnetic fields and  matter-antimatter asymmetries, in the symmetric phase of the early Universe, in the temperature range $100~\mbox{GeV} \le T\le 10~\mbox{TeV}$. Our most important result is that, due to the chiral vortical effect, small overlapping transient fluctuations in the vorticity field in the plasma and temperature of matter degrees of freedom can lead to the generation of strong hypermagnetic fields and matter-antimatter asymmetries, all starting from zero initial values. We show that, either an increase in the amplitudes of the fluctuations of vorticity or temperature, or a decrease in their widths, leads to the production of stronger hypermagnetic fields, and therefore, larger matter-antimatter asymmetries. We have the interesting result that fluctuating vorticity fields are more productive, by many orders of magnitude, as compared to vorticities that are constant in time.


\section{Introduction}
Anomalous transport effects play important roles in particle physics and cosmology, particularly in the early Universe \cite{s1}. One important effect of this kind is the so-called Chiral Vortical Effect (CVE), which refers to the generation of an electric current parallel to the vorticity field in the chiral plasma \cite{av1}. This effect was discovered by Vilenkin who showed that a neutrino current density can result from a rotating black hole \cite{av1}. He obtained the neutrino current density in the direction of the rotation axis as 

\begin{equation}
	J(0)=-\frac{\Omega}{12}T^{2} -\frac{\Omega}{4\pi^{2}}\mu^{2} -\frac{\Omega^{3}}{48\pi^{2}},  
\end{equation}
where $\Omega$ is the angular velocity, $\mu$ is the chiral chemical potential of the neutrino, and $T$ is its temperature.
Thirty years after its discovery, the CVE appeared in the relativistic hydrodynamic equations as an interesting manifestation of anomalies in quantum field theory \cite{son1}.
This effect has attracted much attention and has been investigated extensively in recent years, leading to a deeper understanding of the subject \cite{a1,a11,a2,a3,a4,a5,a6}. In particular, it has been established that in single species chiral plasma in the broken phase at high temperatures, the CVE shows up in the vector current as $\vec{J}_{\mathrm{cv}}=\frac{1}{4\pi^2}(\mu_{R}^2-\mu_{L}^2)
\vec{\Omega}$, and in the axial current as $\vec{J}^{5}_{\mathrm{cv}}=\left[\frac{T^2}{6}+\frac{1}{4\pi^2}\left(\mu_{R}^2+\mu_{L}^2\right)\right]\vec{\Omega}$, where $\mu_{R}$ and $\mu_{L}$ are the right-handed and the left-handed chemical potentials of the species, respectively \cite{av1,son1,a1,a11,a2,a3,a4,a5,a6}\footnote{There are additional contributions of $O(\mu /T)$, where $\mu$ is the chemical potential, which are significant at lower temperatures. See Appendix A for details.}. Interestingly, the term proportional to $T^2$ indicates that there can be an axial current, even if $\mu_{R}=\mu_{L}=0$. 

In this study, we present the correct form of the chiral vortical current in the symmetric phase. Then, we show the prominent effects of the temperature-dependent part of this current in the symmetric phase of the early Universe close to the electroweak phase transition (EWPT). In particular, we show that even very small, but overlapping, transient fluctuations in the vorticity field and temperature of matter degrees of freedom can have important consequences, including the generation of  hypermagnetic field in the absence of initial matter asymmetries. The vorticity fluctuations that we consider are about the zero background value, while the temperature fluctuations are about the finite equilibrium temperature of the plasma. The most important role of the CVE in this context is to produce the magnetic fields, either through the chiralities, or through the temperature fluctuation, the latter of which is the main focus of this work.
Henceforth, we shall refer to transient fluctuations, which we take to be in the form of short pulses, simply as fluctuations.

Another anomalous transport effect is the chiral magnetic effect (CME), which refers to the generation of an electric current parallel to the magnetic field in the imbalanced chiral plasma \cite{avi1,khar1,khar2,khar3}.
It is known that, in single species chiral plasma in the broken phase at high temperatures, the CME appears in the vector current as $\vec{J}_{\mathrm{cm}}=\frac{Q}{4\pi^2}\left(\mu_{R}-\mu_{L}\right)
\vec{B}$, and in the axial current as $\vec{J}_{\mathrm{cm}}^5=\frac{Q}{4\pi^2}\left(\mu_{R}+\mu_{L}\right)
\vec{B}$, where $Q$ is the electric charge \cite{avi1,wang1,khar4,a6} of the species.\footnotemark[1]
Later, we will present the correct form of the chiral magnetic current, in the symmetric phase.
The chiral magnetic current originating from the electroweak Abelian anomaly, and the chiral vortical current are both non-dissipative currents which can strongly affect the generation and the evolution of the magnetic fields and the matter-antimatter asymmetries in the early Universe \cite{sh1,sh2,sh3,Abbaslu1,Elahi-2020}. 


Observations clearly show that our Universe is magnetized on all scales \cite{1,2}.
Various models have been proposed to explain the origin of these magnetic fields \cite{b1,c1,c2,c3,c4,c5,c6,Abbaslu1}, among which the one relying on the electroweak Abelian anomaly has attracted much attention and has been considerably investigated \cite{vb1,sh1,sh2,sh3}. There exists a relationship between the generation and the evolution of the hypermagnetic fields and the fermion number densities in this model, which is due to the chiral coupling of the hypercharge gauge fields to the fermions before the EWPT\cite{c3}.
In case there is a preexisting asymmetry of the right-handed electrons, their number density is almost conserved far from the EWPT, i.e. $T>10\mbox{TeV}$, due to their tiny Yukawa coupling. For lower temperatures, this asymmetry can be converted to the hypermagnetic helicity according to the Abelian anomaly equation, $\partial_{\mu} j_{{e}_R}^{\mu}\sim\vec{E}_{Y}.\vec{B}_{Y}$ \cite{c3,vh1,vb1,sh1,sh2,sh3}.
The anomaly equation shows that, in a reverse process, a strong helical hypermagnetic field can generate the matter-antimatter asymmetries in the Universe, as well \cite{mas,ms1,kh1}.

Another challenge in particle physics and cosmology is the excess of matter over antimatter, with the measured baryon asymmetry of the Universe being of the order of $\eta_{\mathrm{B}}\sim 10^{-10}$ \cite{{bas1},{bas2},{bas3}}. The three Sakharov conditions\footnote{ i- baryon number violation, ii- C and CP violation, iii- a departure from thermal equilibrium.} should be satisfied in any CPT invariant model used to explain this asymmetry from an initially symmetric Universe \cite{Sakharov1}.

In previous studies based on the electroweak Abelian anomalous model, it has been assumed that there is either a significant amount of matter-antimatter asymmetries to produce the hypermagnetic field, or a strong hypermagnetic field to produce the matter-antimatter asymmetries. The most important result of this study is that the matter-antimatter asymmetries and the hypermagnetic field can all be generated simultaneously from zero initial values, by considering the temperature-dependent CVE before the EWPT. 
To obtain this interesting result, we take into account the effects of the temperature-dependent term of the chiral vortical current on the evolution of the hypermagnetic fields and the matter-antimatter asymmetries, by considering simultaneous small  fluctuations, about the background values, in temperature of the right-handed electrons and the vorticity field, close to the EWPT. We also show that fluctuations in the vorticity field are much more productive than vorticity fields that are constant in time. To be more precise, sharp fluctuations yield results comparable to constant vorticities whose amplitudes are many orders of magnitude larger.


As mentioned above, an underlying assumption of this work is the presence of fluctuations in the plasma of the early Universe, containing all elementary particles and gauge fields. A general description of this plasma in the context of hydrodynamics, as an effective field theory, has been presented \cite{Kapusta-2011gt,Jeon-2015dfa}. In such a physical system, hydrodynamic variables naturally fluctuate around their statistical averages. Such stochastic and frequent fluctuations are a commonplace in any plasma, including that of the early Universe. These fluctuations, which may stem from a sum of weakly-correlated random local events, can occur for physical observables such as the number density, velocity, and temperature \cite{Kapusta-2011gt,sen22,Bhattacharyya,Chernodub-2015gxa}. Even though the gauge interactions are strong and their rates are fast in this epoch, the extreme temperatures of the primordial plasma imply that fluctuations could still occur for all species. Moreover, owing to the different Yukawa and gauge couplings, and in particular the chiral nature of the electroweak sector, different chiral fermions could, in principle, experience different fluctuations. All that is required here is that the fluctuations for at least one of the species be non-identical to the rest, at least once. 

Another justification for our hypothesis is the following. Realistic fluctuations of density, temperature and vorticity are usually local in space and time. Meanwhile, in a multi-component plasma local and independent density fluctuations can occur for different matter components. The concurrent occurrence of density, temperature and vorticity fluctuations leads to the possibility of occurrence of non-identical fluctuations for different species of particles. To be concrete, consider a local density fluctuation leading to a local excess of a particle species, e.g. $e_R$, in the central region and rarefaction in the surrounding region, followed immediately by a temperature fluctuation which increases the temperature of the central region. The concurrent occurrence of these two fluctuations leads to a temporary local, and to a much lesser degree global average, increase of temperature of that particle species relative to quasi-equilibrium temperature of the plasma. All that is required now is an overlapping vorticity fluctuation to have the necessary conditions for our model. In this work, we make the simplifying assumption, as is usually done, of considering these fluctuations to be global rather than local. It should be noted that the amplitude and duration of these fluctuations are usually small enough such that the equilibrium energy density, pressure and entropy of the system can be defined. Nevertheless, as we shall show, even very small and brief fluctuations can trigger the mechanism that we propose here and affect significantly the dynamical evolution of the system, as is formulated within the equations of anomalous magnetohydrodynamics (AMHD).

This paper is organized as follows: In Sec.\ \ref{x1}, the anomalous magnetohydrodynamics equations and the evolution equations for the matter-antimatter asymmetries are derived in the expanding Universe. In Sec.\ \ref{x4}, the set of coupled differential equations are solved numerically. In Sec.\ \ref{x5}, the results are summarized and the conclusion is presented.

\section{Evolution Equations}\label{x1}
\subsection{Anomalous Magnetohydrodynamics Equations}
In this section, the anomalous magnetohydrodynamics (AMHD) equations are obtained in the Landau-Lifshitz frame in the symmetric phase of the expanding Universe. 
Taking the CVE and the CME into account, the Maxwell's equations for the hypercharge-neutral plasma in the expanding Universe are given as \cite{son1,{6},{dettmann},{Subramanian}, Abbaslu1,Yamamoto16,Anand:2017,Landsteiner-2016,Landsteiner-2012kdw,Neiman-2010zi}
\begin{equation}\label{eq1}
\frac{1}{R}\vec{\nabla} .\vec{E}_{Y}=0,\qquad\qquad\frac{1}{R}\vec{\nabla}.\vec{B}_{Y}=0,
\end{equation}
\begin{equation}\label{eq2}
\frac{1}{R}\vec{\nabla}\times\vec{ E}_{Y}+\left(\frac{\partial \vec{B}_{Y}}{\partial t}+2H\vec{B}_{Y}\right)=0,
\end{equation}
\begin{equation}\label{eq3}
\begin{split}
\frac{1}{R}\vec{\nabla}\times\vec{B}_{Y}-&\left(\frac{\partial \vec{E}_{Y}}{\partial t}+2H\vec{E}_{Y}\right)=\vec{J}\\&=\vec{J}_{\mathrm{Ohm}}+\vec{J}_{\mathrm{cv}}+\vec{J}_{\mathrm{cm}},
\end{split}
\end{equation}
\begin{equation}\label{eq2.3}
\vec{J}_{\mathrm{Ohm}}=\sigma\left(\vec{E}_{Y}+\vec{v}\times\vec{B}_{Y}\right),
\end{equation}
\begin{equation}\label{eq3.1}
\vec{J}_{\mathrm{cv}}=c_{\mathrm{v}}\vec{\omega},
\end{equation}
\begin{equation}\label{eq3.2}
\vec{J}_{\mathrm{cm}}=c_{\mathrm{B}}\vec{B}_{Y},
\end{equation}
where $R$ is the scale factor, $H=\dot{R}/R$ is the Hubble parameter, $\sigma$ is the electrical hyperconductivity of the plasma, and $\vec{v}$ and $\vec{\omega}=\frac{1}{R}\vec{\nabla}\times\vec{v}$ are the bulk velocity and vorticity of the plasma, respectively. Furthermore, the chiral vorticity and helicity coefficients $c_{\mathrm{v}}$ and $c_{\mathrm{B}}$ are as follows (see Appendix A)\footnote{The temperature-independent parts of these coefficients were presented in \cite{Abbaslu1}. There are additional terms of $O(\mu/T)$ which, as we shall show explicitly in Sec.\ \ref{x4}, are negligible within the initial conditions and results of our model (see Appendix A).},
\begin{equation}\label{eq22}
\begin{split}
c_{\mathrm{v}}(t)=&\sum_{i=1}^{n_{G}}\Big[\frac{g'}{48}\Big(-Y_{R}T_{R_{i}}^{2}+Y_{L}T_{L_{i}}^{2}N_{w}-Y_{d_{R}}T_{d_{R_{i}}}^{2}N_{c}-Y_{u_{R}}T_{u_{R_{i}}}^{2}N_{c}+Y_{Q}T_{Q_{i}}^{2}N_{c}N_{w}\Big)\\&+\frac{{g'}}{16\pi^{2}}\Big(-Y_{R}\mu_{R_{i}}^{2}+Y_{L}\mu_{L_{i}}^{2}N_{w}-Y_{d_{R}}\mu_{d_{R_{i}}}^{2}N_{c}-Y_{u_{R}}\mu_{u_{R_{i}}}^{2}N_{c}+Y_{Q}\mu_{Q_{i}}^{2}N_{c}N_{w}\Big)\Big],	
\end{split}
\end{equation}
\begin{equation}\label{eqc_B} 
\begin{split}
c_{\mathrm{B}}(t)=
&-\frac{g'^{2}}{8\pi^{2}}\sum_{i=1}^{n_{G}}\Big[-\Big(\frac{1}{2}\Big)Y_{R}^{2}\mu_{R_{i}}-\Big(\frac{-1}{2}\Big)Y_{L}^{2}\mu_{L_{i}}N_{w}-\Big(\frac{1}{2}\Big)Y_{d_{R}}^{2}\mu_{d_{R_{i}}}N_{c}-\Big(\frac{1}{2}\Big)Y_{u_{R}}^{2}\mu_{u_{R_{i}}}N_{c}\\&-\Big(\frac{-1}{2}\Big)Y_{Q}^{2}\mu_{Q_{i}}N_{c}N_{w}\Big], 
\end{split}
\end{equation}
where $n_{G}$ is the number of generations, and $N_{c}=3$ and $N_{w}=2$ are the ranks of the non-Abelian SU$(3)$ and SU$(2)$ gauge groups, respectively. Moreover, $\mu_{L_i}$($\mu_{R_i}$), $\mu_{Q_i}$, and $\mu_{{u_R}_i}$ ($\mu_{{d_R}_i}$) are the common chemical potentials of left-handed (right-handed) leptons, left-handed quarks with different colors, and up (down) right-handed quarks with different colors, respectively. Furthermore, \lq{\textit{i}}\rq\ is the generation index, and the relevant hypercharges are
\begin{equation}\label{eqds}
\begin{split}
	&Y_{L}=-1, \quad Y_{R}=-2,\\& Y_{Q}=\frac{1}{3}, \quad Y_{u_{R}}=\frac{4}{3}, \quad Y_{d_{R}}=-\frac{2}{3}.
\end{split}	
\end{equation}
After substituting the hypercharges in Eqs.\ (\ref{eq22}) and (\ref{eqc_B}), we obtain
\begin{equation}\label{eq24}
\begin{split}
c_{\mathrm{v}}(t)=&\sum_{i=1}^{n_{G}}\Big[\frac{g'}{24}\left(T_{R_{i}}^{2}-T_{L_{i}}^{2}+T_{d_{R_{i}}}^{2}-2T_{u_{R_{i}}}^{2}+T_{Q_{i}}^{2}\right)+\frac{{g'}}{8\pi^{2}}\Big(\mu_{R_{i}}^{2}-\mu_{L_{i}}^{2}+\mu_{d_{R_{i}}}^{2}-2\mu_{u_{R_{i}}}^{2}+\mu_{Q_{i}}^{2}\Big)\Big],
\end{split}	
\end{equation}
\begin{equation}\label{eq26}
\begin{split} 
&c_{\mathrm{B}}(t)=\frac{-g'^{2}}{8\pi^{2}} \sum_{i=1}^{n_{G}}\left[-2\mu_{R_{i}}+\mu_{L_{i}}-\frac{2}{3}\mu_{d_{R_{i}}}-\frac{8}{3}\mu_{u_{R_{i}}}+\frac{1}{3}\mu_{Q_{i}}\right]. 
\end{split}
\end{equation}

Let us make the same assumptions as in our previous studies, and simplify $c_{\mathrm{v}}$ and $c_{\mathrm{B}}$ accordingly \cite{Abbaslu1,sh2}.
We assume that all quark Yukawa processes
are in equilibrium and, because of the flavor mixing in the quark sector, all up or down quarks belonging to different generations with distinct handedness have the same chemical potential \cite{sh2,17}. For simplicity, we also assume that the Higgs asymmetry is zero and obtain \cite{27,sh2}
\begin{equation}\label{eq29}
\mu_{u_{R}}=\mu_{d_{R}}=\mu_{Q}.
\end{equation}
Furthermore, we assume that only the contributions of the baryonic and the first-generation leptonic chemical potentials to $c_{\mathrm{v}}$ and $c_{\mathrm{B}}$ are significant. As for the temperature fluctuations, it suffices to consider fluctuations in only one of the matter components, which we take to be $e_{R}$. Using Eq.\ (\ref{eq29}) and the aforementioned assumptions, we simplify Eqs.\ (\ref{eq24}) and (\ref{eq26}) to obtain 
\begin{equation}\label{eq321}
c_{\mathrm{v}}(t)=\frac{g'}{24}\left(\Delta T^{2}\right)+\frac{g'}{8\pi^{2}}\left(\mu_{e_{R}}^{2}-\mu_{e_{L}}^{2}\right),	
\end{equation}
\begin{equation}\label{eq33}
c_{\mathrm{B}}(t)=-\frac{g'^{2}}{8\pi^{2}}\left(-2\mu_{e_{R}}+\mu_{e_{L}}-\frac{3}{4}\mu_{\mathrm{B}}\right), 
\end{equation}




where $\mu_{\mathrm{B}}=12\mu_{Q}$, and $\Delta T^{2}=T_{e_{R}}^{2}- T^2$ is the temperature fluctuation, and $T$ is the equilibrium temperature of the thermal bath, which includes all other components of the plasma. We set ${\Delta T}^{2}=T^{2} \beta [x(T)]$, where $\beta [x(T)]$ is an arbitrary profile function to be specified later, and $x(T)=t(T)/t_\mathrm{EW}=\left(T_\mathrm{EW}/T\right)^{2}$ is given by the Friedmann law.

In the Landau-Lifshitz frame, the continuity and Navier-Stokes equations are given as follows \cite{Yamamoto16,Anand:2017,Landsteiner-2016,Abbaslu1}

\begin{equation}\label{eq7}
	\frac{\partial \rho}{\partial t}+\frac{1}{R}\vec{\nabla}.\left[\left(\rho+p\right)\vec{v}\right]+3H\left(\rho+p\right)=0,
	\end{equation}
\begin{equation}\label{eq4.1}
	\begin{split}
	&\left[\frac{\partial}{\partial t}+\frac{1}{R}\left(\vec{v}.\vec{\nabla}\right)+H\right]\vec{v}+\frac{\vec{v}}{\rho+p}\frac{\partial p}{\partial t}=-\frac{1}{R}\frac{\vec{\nabla} p}{\rho+p}+\frac{\vec{J}\times\vec{B}_{Y}}{\rho+p}+\frac{\nu}{{R}^{2}}\left[\nabla^{2}\vec{v}+\frac{1}{3}\vec{\nabla}\left(\vec{\nabla}.\vec{v}\right)\right],
	\end{split}
	\end{equation}	
where $\nu$ is the kinematic viscosity, and $\rho$ and $p$ 
are the energy density and the pressure of the plasma, respectively. 
Combining the fluid incompressibility condition in the lab frame, $\partial_{t}\rho+3H\left(\rho+p\right)=0$ or equivalently $H\vec{v}+\vec{v}\partial_{t} p/\left(\rho+p\right)=0$, with Eq.\ (\ref{eq7}) leads to the condition $\vec{\nabla}.\vec{v}=0$ \cite{Abbaslu1,555,Anand:2017}.

In the following, we choose a simple monochromatic Chern-Simons configuration for the hypermagnetic field $\vec{B}_Y=(1/R)\vec{\nabla} \times \vec{A}_Y$, and the velocity field $\vec{v}=(1/R)\vec{\nabla} \times \vec{S}$ \cite{{sh2},Abbaslu1}. To do this, we choose $\vec{A}_{Y}=\gamma(t)\left(\cos kz , \sin kz, 0\right)$, and $\vec{S}=r(t)\left(\cos kz , \sin kz, 0\right)$, for their corresponding vector potentials \cite{69,70,76}.
Note that we have chosen a fully helical form with the negative helicity for both, the reason for which will be stated later.
Let us now obtain the evolution equation for the velocity field. Neglecting the displacement current in the lab frame in Eq.\ (\ref{eq3}), the total current becomes $\vec{J}=(1/R)\vec{\nabla}\times\vec{B}_{Y}$, and as a result, $\vec{J}\times\vec{B}_{Y}$ vanishes in Eq.\ (\ref{eq4.1}).
Then, using $\vec{\nabla} .\vec{v}=0$ and $H\vec{v}+\vec{v}\partial_{t}p/(\rho+p)=0$, as stated earlier, and neglecting the gradient terms in Eq.\ (\ref{eq4.1}), the evolution equation for the velocity field becomes \footnote{The term $(\vec{v}.\vec{\nabla})\vec{v}$ is neglected because of being next to leading order. Furthermore, the magnetic pressure $B^{2}/8\pi$ is negligible compared to the fluid (radiation) pressure $p$ which is homogeneous and isotropic. Indeed, the maximum value of their ratio at the onset of the EWPT is $B^{2}/8\pi p\approx 10^{-7} \ll 1$. Therefore, to a good approximation, the homogenity and isotropy conditions remain valid and the pressure variations in the fluid $\vec{\nabla} p$ can be neglected\cite{Pavlotic}.}
 
 \begin{equation}\label{eq19}
 \frac{\partial \vec{v}}{\partial t}=-\nu{k^{\prime}}^{2}\vec{v},
 \end{equation}
where the kinematic viscosity $\nu\simeq1/(5\alpha_{Y}^{2}T)$ \cite{{41},{banerjee}}. 
Neglecting the displacement current in the lab frame and using the aforementioned configurations, the hyperelectric field and the evolution equation for the hypermagnetic field are obtained, as follows:
\begin{equation}\label{eq16}
\vec{E}_{Y}=-\frac{k^{\prime}}{\sigma }\vec{B}_{Y}+\frac{c_{\mathrm{v}}}{\sigma }k^{\prime}\vec{v}-\frac{c_{\mathrm{B}}}{\sigma}\vec{B}_{Y},
\end{equation}
\begin{equation}\label{eq17}
\begin{split}
\frac{dB_{Y}(t)}{dt}=&\left[-\frac{1}{ t}-\frac{{k^{\prime}}^{2}}{\sigma } -\frac{c_{\mathrm{B}}k^{\prime}}{\sigma } \right]B_{Y}(t)+\frac{c_{\mathrm{v}}}{\sigma}{k^{\prime}}^{2}\langle\vec{v}(t).\hat{B}_{Y}(t)\rangle,
\end{split}
\end{equation}
where $\vec{\omega}=-k^{\prime}\vec{v}$, $\sigma=100T$,  angle brackets denote spatial average, and $k^{\prime}=k/R=kT$. The latter shows the increase of the hypermagnetic length scale, due to the expansion of the Universe. 
Note that with the choice of vector potentials for $\vec{v}(t)$ and  $\vec{B}_{Y}(t)$, the advection term $\vec{v}\times\vec{B}_{Y}$ has been set to zero in the above, and we have the following simplification: $\langle\vec{v}(t).\hat{B}_{Y}(t)\rangle \rightarrow v(t)$. In the next subsection we obtain the evolution equations for the matter-antimatter asymmetries.

\subsection{Evolution equations for the matter-antimatter asymmetries}\label{x3}
Before the EWPT, the gauge fields of $\textrm{U}_\textrm{Y}(1)$ couple to the fermions chirally, in contrast to those of $\textrm{U}_\textrm{em}(1)$ in the broken phase, leading to the non-conservation of the matter currents. This shows up in the Abelian anomaly equations \cite{{26}}, which, for the first-generation leptons, are
\begin{equation}\label{anomaly1}
\begin{split}
&\nabla_{\mu} j_{{e}_R}^{\mu}=-\frac{1}{4}(Y_{R}^{2})\frac{g'^{2}}{16 \pi^2}Y_{\mu\nu}\tilde{Y}^{\mu\nu}=\frac{g'^{2}}{4\pi^{2}}\vec{E}_{Y}.\vec{B}_{Y},\\
&\nabla_{\mu} j_{{e}_L}^{\mu}=\frac{1}{4}(Y_{L}^{2})\frac{g'^{2}}{16\pi^2}Y_{\mu\nu}\tilde{Y}^{\mu\nu}=-\frac{g'^{2}}{16\pi^{2}}\vec{E}_{Y}.\vec{B}_{Y},
 \end{split}
\end{equation}
where $\nabla_{\mu}$ is the covariant derivative with respect to Friedmann-Robertson-Walker (FRW) metric $ds^{2}=dt^{2}-R^{2}(t)\delta_{ij}dx^{i}dx^{j}$, $t$ is the physical time, and $x^{i}$s are the comoving coordinates. Integrating the above equations over all space and considering the perturbative chirality flip reactions for the leptons, we obtain (see Refs.\cite{sh1,sh2,28,Abbaslu1} and Appendix B for details),  
\begin{equation}\label{anomaly2}
\begin{split}
	\frac{d\eta_{{e}_{R}}}{dt}=&\frac{g'^{2}}{4\pi^{2} s}\langle\vec{E}_{Y}.\vec{B}_{Y}\rangle+\left(\frac{\Gamma_{0}}{t_{EW}}\right)\left(\frac{1-x}{\sqrt{x}}\right)\left(\eta_{e_{L}}-\eta_{e_{R}}\right)\\&-\frac{d}{dt}\Big[\frac{g^{\prime}\mu_{e_{R}}}{4\pi^{2}s}\langle\vec{v}.\vec{B}_{Y}\rangle\Big]-\frac{d}{dt}\Big[\big(\frac{\mu_{e_{R}}^{2}}{8\pi^{2}s}+\frac{T^{2}}{24s}\big)\langle\vec{v}.\vec{\omega}\rangle\Big]+\frac{d}{dt}\Big[\frac{2\sigma_{e_{R}}}{g^{\prime}Y_{R}s}\langle\vec{v}.\vec{E}_{Y}\rangle\Big]\\
	\frac{d\eta_{\nu_{e}^{L}}}{dt}=&\frac{d\eta_{{e}_{L}}}{dt}=-\frac{g'^{2}}{16\pi^{2} s}\langle\vec{E}_{Y}.\vec{B}_{Y}\rangle+\left(\frac{\Gamma_{0}}{2t_{EW}}\right)\left(\frac{1-x}{\sqrt{x}}\right)\left(\eta_{e_{R}}-\eta_{e_{L}}\right)\\&+\frac{d}{dt}\Big[\frac{g^{\prime}\mu_{e_{L}}}{8\pi^{2}s}\langle\vec{v}.\vec{B}_{Y}\rangle\Big]+\frac{d}{dt}\Big[\big(\frac{\mu_{e_{L}}^{2}}{8\pi^{2}s}+\frac{T^{2}}{24s}\big)\langle\vec{v}.\vec{\omega}\rangle\Big]+\frac{d}{dt}\Big[\frac{2\sigma_{e_{L}}}{g^{\prime}Y_{L}s}\langle\vec{v}.\vec{E}_{Y}\rangle\Big].
	\end{split}
\end{equation}
 In the equations above, $\eta_{f}=\left(n_{f}/s\right)$ with $f=e_{R},e_{L},\nu_{e}^{L}$ is the fermion asymmetry and $n_{f}$ is the charge density of the $f$th species of fermion, $s=2\pi^{2}g^{*}T^{3}/45$ is the entropy density and $g^{*}=106.75$ is the effective number of relativistic degrees of freedom, $x=\left(t/t_\mathrm{EW}\right)=\left(T_\mathrm{EW}/T\right)^{2}$  is given by the Friedmann law, $\Gamma_{0}=121$, $t_\mathrm{EW}=\left(M_{0}/2T_\mathrm{EW}^{2}\right)$, and $M_{0}=\left(M_\mathrm{Pl}/1.66\sqrt{g^{*}}\right)$, where $M_\mathrm{Pl}$ is the Plank mass. Furthermore, the term $\frac{\Gamma_{0}}{t_\mathrm{EW}}\left(\frac{1-x}{\sqrt{x}}\right)$ appearing in the equations is the chirality flip rate of the right-handed electrons.  
In a similar manner, the evolution equation for the baryon asymmetry can also be obtained as (see Appendix B)
	\begin{equation}
	\frac{d\eta_{\mathrm{B}}}{dt}=\frac{3g'^{2}}{8\pi^{2} s}\langle\vec{E}_{Y}.\vec{B}_{Y}\rangle+\frac{d}{dt}\left[\frac{2}{sg'}(\frac{\sigma_{d_R}}{Y_{d_R}}+\frac{\sigma_{u_R}}{Y_{u_R}}+2\frac{\sigma_{Q}}{Y_{Q}})\langle\vec{v}.\vec{E}_{Y}\rangle\right].
	\end{equation}		
Using $\mu_{f}=(6s/T^2)\eta_f$  with Eq.\ (\ref{eq16}) we obtain
\begin{equation}\label{eq44}
\begin{split}
\langle\vec{E}_{Y}.\vec{B}_{Y}\rangle=&\frac{B_{Y}^{2}(t)}{100} \left[-\frac{k^{\prime}}{T}-\frac{6sg'^{2}}{4\pi^{2}T^3}\left(\eta_{e_{R}}-\frac{\eta_{e_{L}}}{2}+\frac{3}{8}\eta_{\mathrm{B}}
\right)\right]\\&+\left[\frac{g'}{24}\beta [x(T)]+\frac{36s^2g'}{8\pi^{2}T^6}\left(\eta_{e_{R}}^{2}-\eta_{e_{L}}^{2}\right)\right]\frac{k^{\prime}T }{100}\langle\vec{v}(t).\vec{B}_{Y}(t)\rangle.
\end{split}
\end{equation}
With the helical configurations chosen, $\langle\vec{v}(t).\vec{B}_{Y}(t)\rangle \rightarrow v(t)B_Y(t)$. Using $1\mbox{Gauss}\simeq2\times10^{-20} \mbox{GeV}^{2}$, and setting the kinematic viscosity $\nu$ to zero for simplicity, we obtain the complete set of evolution equations for the matter-antimatter asymmetries and the amplitudes of the hypermagnetic and velocity fields as
	\begin{equation}\label{eq47}
	\begin{split}
	\frac{d\eta_{e_R}}{dx}&=\frac{1}{\lambda_{R}(x)}\Bigg[\Lambda_{R}(x)+\left[-C_{1}-C_{2} \eta_{T}(x)
	\right]\left(\frac{B_{Y}(x)}{10^{20}G}\right)^{2}x^{3/2}\\&+\left[C_{3}\beta(x)+C_{4} \Delta \eta^{2}(x)\right]v(x)\left(\frac{B_{Y}(x)}{10^{20}G}\right)\sqrt{x}\Bigg]-\Gamma_{0}\frac{1-x}{\sqrt{x}}\left[\eta_{e_R}(x)-\eta_{e_L}(x)\right],
	\end{split}
	\end{equation}
	\begin{equation}\label{eq48}
	\begin{split}
	\frac{d\eta_{e_L}}{dx}&=\frac{1}{\lambda_{L}(x)}\Bigg[\Lambda_{L}(x)-\frac{1}{4}\left[-C_{1}-C_{2}\eta_{T}(x)
	\right]\left(\frac{B_{Y}(x)}{10^{20}G}\right)^{2}x^{3/2}\\&-\frac{1}{4}\left[C_{3}\beta(x)+C_{4} \Delta \eta^{2}(x)\right]v(x)\left(\frac{B_{Y}(x)}{10^{20}G}\right)\sqrt{x}\Bigg]+\Gamma_{0}\frac{1-x}{2\sqrt{x}}\left[\eta_{e_R}(x)-\eta_{e_L}(x)\right],
	\end{split}
	\end{equation}
	\begin{equation}\label{eq49} 
	\begin{split}
	\frac{dB_{Y}}{dx}&=\frac{1}{\sqrt{x}}\left[-C_{5} -C_{6}\eta_{T}(x)
	\right]B_{Y}(x)-\frac{1}{x}B_{Y}(x)+\left[C_{7}\beta(x)+C_{8}\Delta \eta^{2}(x)\right]\frac{v(x)}{x^{3/2}} ,
	\end{split}
	\end{equation}	
	\begin{equation}\label{eq54}
\begin{split}
\frac{d\eta_{\mathrm{B}}}{dx}&=\frac{3}{2}\Bigg[-C_{1}-C_{2}\eta_{T}(x)
	\left(\frac{B_{Y}(x)}{10^{20}G}\right)^{2}x^{3/2}+\left[C_{3}\beta(x)+C_{4} \Delta \eta^{2}(x)\right]v(x)\left(\frac{B_{Y}(x)}{10^{20}G}\right)\sqrt{x}\Bigg]\\&+\frac{81x}{100Mg^{\prime}10^{20}  G}\langle\vec{E}_{Y}(x).\partial_ x\vec{v}(x)\rangle
\end{split}
\end{equation}
		\begin{equation}
		\begin{split}
		&\Delta \eta^{2}(x)= \eta_{e_R}^{2}(x)-\eta_{e_L}^{2}(x),\\&
		\eta_{T}(x)=\eta_{e_R}(x)-\frac{\eta_{e_L}(x)}{2}+\frac{3}{8}\eta_{\mathrm{B}(x)},\\
		&\lambda_{R}(x)=1-\frac{6g^{\prime}Y_{R}}{8\pi^{2}}\frac{\langle\vec{v}(x).\vec{B}_{Y}(x)\rangle}{10^{20}G}\frac{x}{5000}-\frac{36M}{4\pi^{2}}\eta_{e_R}(x)kv^{2}(x),\\&
		\lambda_{L}(x)=1+\frac{6g^{\prime}Y_{L}}{8\pi^{2}}\frac{\langle\vec{v}(x).\vec{B}_{Y}(x)\rangle}{10^{20}G}\frac{x}{5000}+\frac{36M}{4\pi^{2}}\eta_{e_L}(x)kv^{2}(x),\\
		\Lambda_{R}(x)=&\frac{6g^{\prime}Y_{R}}{8\pi^{2}}\frac{x}{5000}\frac{\eta_{e_R}(x)}{10^{20}G}\Big[\langle\vec{v}(x).\partial_{x}\vec{B}_{Y}(x)\rangle+\langle\frac{\vec{v}(x).\vec{B}_{Y}(x)}{x}\rangle+\langle\vec{B}_{Y}(x).\partial_{x}\vec{v}(x)\rangle\Big]\\&+\Big[\frac{36M}{4\pi^{2}}\eta_{e_R}^2(x)+\frac{1}{12M}\Big]k\vec{v}(x).\partial_{x}\vec{v}(x)+\frac{x}{25Mg^{\prime}Y_{R}10^{20}  G}\langle\vec{E}_{Y}(x).\partial_{x}\vec{v}(x)\rangle,\\
		\Lambda_{L}(x)=&-\frac{6g^{\prime}Y_{L}}{8\pi^{2}}\frac{x}{5000}\frac{\eta_{e_L}(x)}{10^{2}G}\Big[\langle\vec{v}(x).\partial_{x}\vec{B}_{Y}(x)\rangle+\langle\frac{\vec{v}(x).\vec{B}_{Y}(x)}{x}\rangle+\langle\vec{B}_{Y}(x).\partial_{x}\vec{v}(x)\rangle\Big]\\&-\Big[\frac{36M}{4\pi^{2}}\eta_{e_L}^{2}(x)+\frac{1}{12M}\Big]k\vec{v}(x).\partial_{x}\vec{v}(x)+\frac{x}{25Mg^{\prime}Y_{L}10^{20}  G}\langle\vec{E}_{Y}(x).\partial_{x}\vec{v}(x)\rangle,\\&
		\end{split}
		\end{equation}
where $M=2\pi^{2}g^{*}/45$, and the coefficients $C_{i}, i=1,...,8$, are 
	\begin{equation}\label{eq51}
	\begin{split}
	& C_{1}=0.00096\left(\frac{k}{10^{-7}}\right)\alpha_{Y},\\&
	 C_{2}=865688 \alpha_{Y}^{2},\\&
	  C_{3}=0.71488\left(\frac{k}{10^{-7}}\right)\alpha_{Y}^{3/2},\\&
	  C_{4}= 17152.7\left(\frac{k}{10^{-7}}\right)\alpha_{Y}^{3/2},\\&
	   C_{5}= 0.356\left(\frac{k}{10^{-7}}\right)^{2},\\&
	   C_{6}=3.18373 \times10^{8} \alpha_{Y} \left(\frac{k}{10^{-7}}\right), \\&
	C_{7}=262.9\times 10^{20} \sqrt{\alpha_{Y}}\left(\frac{k}{10^{-7}}\right)^{2},\\& 
	C_{8}=63\times 10^{25} \sqrt{\alpha_{Y}}\left(\frac{k}{10^{-7}}\right)^{2},\\&
	\end{split}
	\end{equation}
 and $\alpha_{Y}=g'^{2}/4\pi\simeq0.01$ is the fine-structure constant for the $\textrm{U}_\textrm{Y}(1)$.
We now choose the profile of temperature fluctuation $\beta [x(T)] = \Delta T^2/T^2$, as defined in Eq.\ (\ref{eq321}) and the paragraph below it, to be a Gaussian function of $x$:
 \begin{equation}\label{Tprofile}
  \beta(x)=\frac{\beta_{0}}{b\sqrt{2\pi}}\exp\left[-\frac{(x-x_{0})^{2}}{2b^{2}}\right], 
 \end{equation}
where $\beta_{0}$ is the amplitude multiplying the normalized Gaussian distribution, and $x=\left(t/t_{EW}\right)=\left(T_{EW}/T\right)^{2}$, as defined before. 
The profile of the vorticity fluctuation must have an overlap with that of temperature fluctuation, in order to produce any effect. For simplicity, we choose the two profiles to be identical. That is, 
\begin{equation}\label{vprofile}
\omega (x)=k^{\prime} v(x)=\frac{k^{\prime}v_{0}}{b\sqrt{2\pi}}\exp\left[-\frac{(x-x_{0})^{2}}{2b^{2}}\right],
\end{equation}
where $v_{0}$ is the amplitude of the velocity fluctuation.
We should mention that the occurrence of any fluctuation in a plasma in a quasi-equilibrium state would normally trigger a restoring response originating from dissipative effects, such as viscous effects. Here, for simplicity we assume that the combined results of the original fluctuations and  the ensuing dissipative effects have the profiles given by Eqs.\ (\ref{Tprofile},\ref{vprofile}).

The majority of analysis presented in the following section is for a single pulse for temperature and vorticity, as stated above, which, as we shall show, produce matter-antimatter asymmetries and helical hypermagnetic field. However, at the end of the next section we present the results for two sets of successive pulses, the latter one having the same temperature profile but with negative amplitude, showing that the first pulse is the main determinant of the outcome.

\section{Numerical Solution}\label{x4}
In this section, we obtain the numerical solutions of the evolution equations. As mentioned earlier, we investigate the effects of the temperature fluctuations of right-handed electrons, in the presence of vorticity, on the generation and evolution of the hypermagnetic field and the matter-antimatter asymmetries, in the temperature range $100 \mbox{ GeV}\leq T \leq 10\mbox{ TeV}$. We consider the temperature fluctuations as small Gaussian distributions in $x$, as shown in Eq.\ (\ref{Tprofile}), that occur close to the EWPT. As for the vorticity field, we consider small fluctuations, whose profiles, as shown in Eq.\ (\ref{vprofile}), coincide with those of the temperature fluctuations. We also investigate the cases with constant vorticity fields for comparison. As we shall show, the former is much more interesting and will be the focus of our work, since it is not only physically more realistic, but also could yield orders of magnitude larger results for the asymmetries and the hypermagnetic field. In the following, we solve the evolution equations by considering the comoving wave number as $k=10^{-7}$, and setting the initial values of the hypermagnetic field amplitude and the matter-antimatter asymmetries to zero, i.e.\ $B_{Y}^{(0)}=0$, and $\eta_{e_R}^{(0)}=\eta_{e_L}^{(0)}=\eta_{B}^{(0)}=0$.

\begin{figure*}[!ht]
	\subfigure[]{\label{fig:figure:1}
		\includegraphics[width=.45\textwidth]{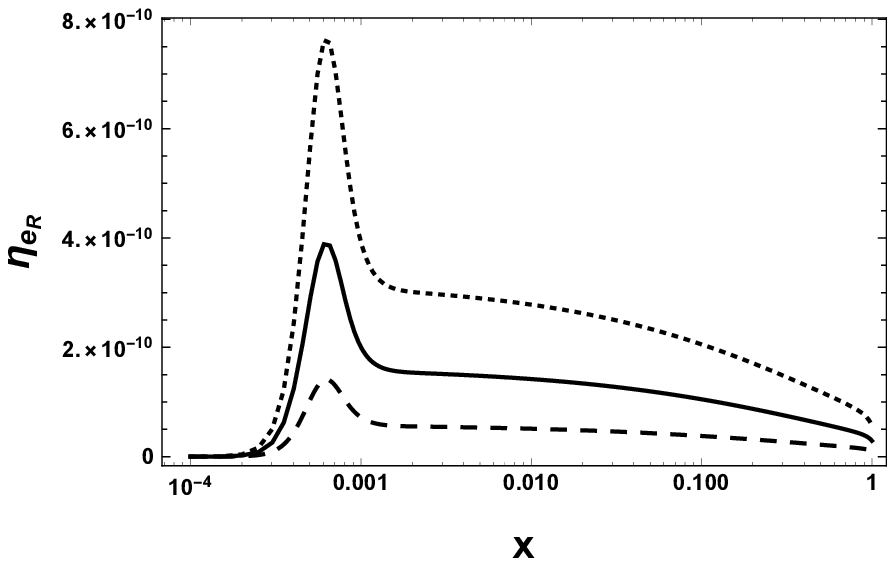}}
	\hspace{8mm}
	\subfigure[]{\label{fig:figure:2}
		\includegraphics[width=.45\textwidth]{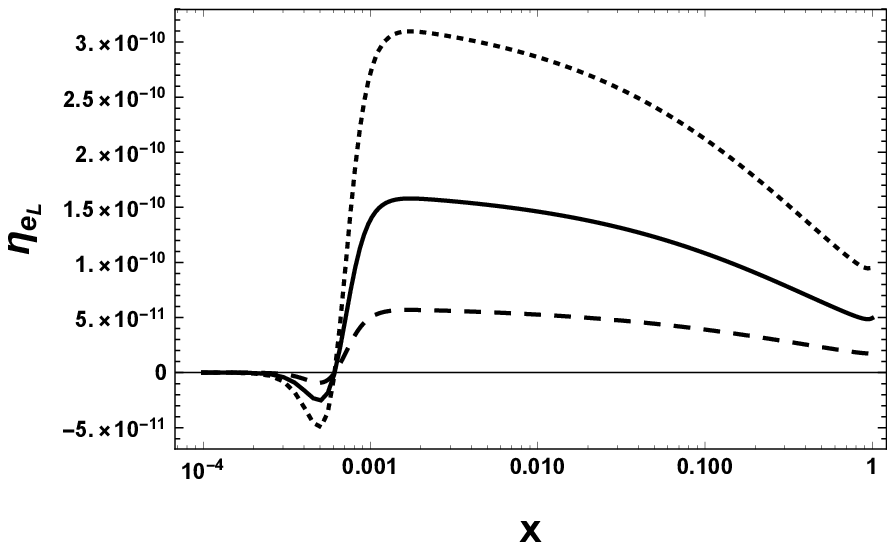}}
	\hspace{8mm}
	\subfigure[]{\label{fig:figure:11}
		\includegraphics[width=.45\textwidth]{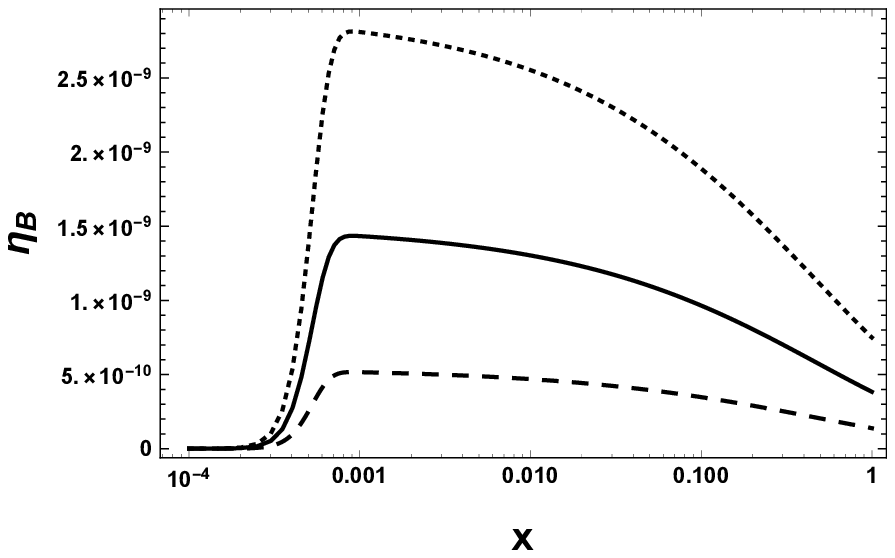}}
	\hspace{8mm}
	\subfigure[]{\label{fig:figure:21} 
		\includegraphics[width=.45\textwidth]{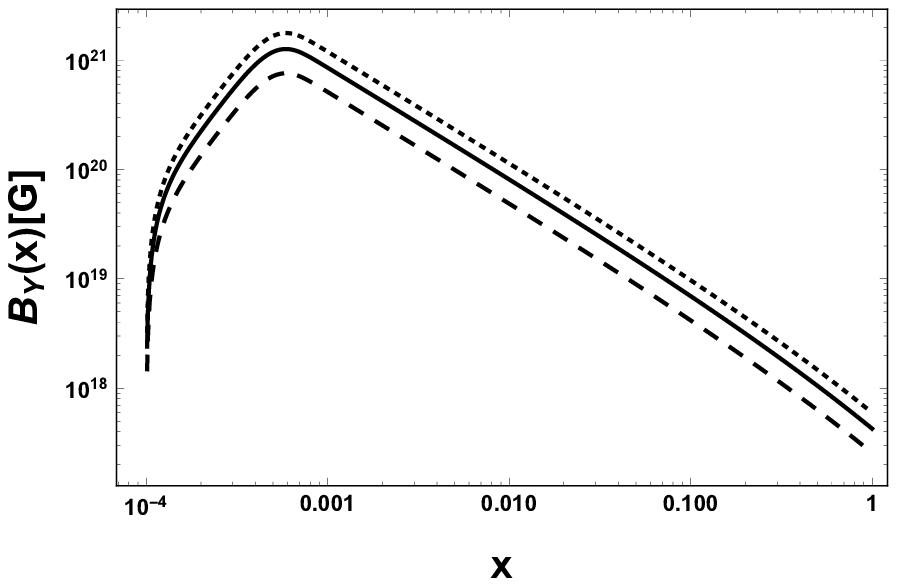}}
	\hspace{8mm}
	\caption{\footnotesize Time plots of: (a) the right-handed electron asymmetry $\eta_{e_{R}}$, (b) the left-handed electron asymmetry $\eta_{e_{L}}$, (c) the baryon asymmetry $\eta_{B}$, and (d) the hypermagnetic field amplitude $B_{Y}$, for various values of the amplitude of temperature fluctuation of $e_{R}$. The initial conditions are: $k=10^{-7}$, $B_{Y}^{(0)}=0$, $\eta_{e_R}^{(0)}=\eta_{e_L}^{(0)}=\eta_{B}^{(0)}=0$, $v_0=10^{-5}$, $b=2\times10^{-4}$, and $x_{0}=45\times10^{-5}$. The dashed line is for $\beta_{0}=3\times10^{-4}$, the solid line is for $\beta_{0}=5\times10^{-4}$, and the dotted line is for $\beta_{0}=7\times10^{-4}$.}
	\label{fig1}
\end{figure*}

For our first case, we solve the coupled differential equations with the initial conditions, $v_0=10^{-5}$, $b=2\times10^{-4}$, and $x_{0}=45\times10^{-5}$, for various values of the amplitude of temperature fluctuations $\beta_{0}$, and present the results in Fig.\ \ref{fig1}. As can be seen in the figure, the simultaneous occurrence of small vorticity fluctuation and temperature fluctuation for the right handed electrons leads to the generation of strong hypermagnetic fields which then produce the matter-antimatter asymmetries, all starting from zero initial values. It can be seen that by increasing the amplitude of the temperature fluctuation, the maximum and the final values of the hypermagnetic field amplitude, as well as the matter-antimatter asymmetries, increase. We have found that in our model signs of the matter-antimatter asymmetries produced and the helicity of hypermagnetic and vorticity fields, are always opposite. This is a manifestation of the generalized charge conservation as stated in Eq.\ (\ref{eq4aq3wsaw1efs}) of Appendix B.\footnote{In fact, the second and fourth terms in Eq.\ (\ref{eq4aq3wsaw1efs}) are negligible in this study.} The Chern-Simons configuration that we have chosen for the hypermagnetic and vorticity fields has negative helicity. Figure \ref{fig1} also shows that the hypermagnetic field amplitude grows to its maximum value of about $10^{21}$G, then decreases due to the expansion of the Universe. We have also investigated the effects of changing the amplitude of the vorticity fluctuation, and have found similar results. 


%

\begin{figure*}[!ht]
	\subfigure[]{\label{fig:figure:111}
		\includegraphics[width=.45\textwidth]{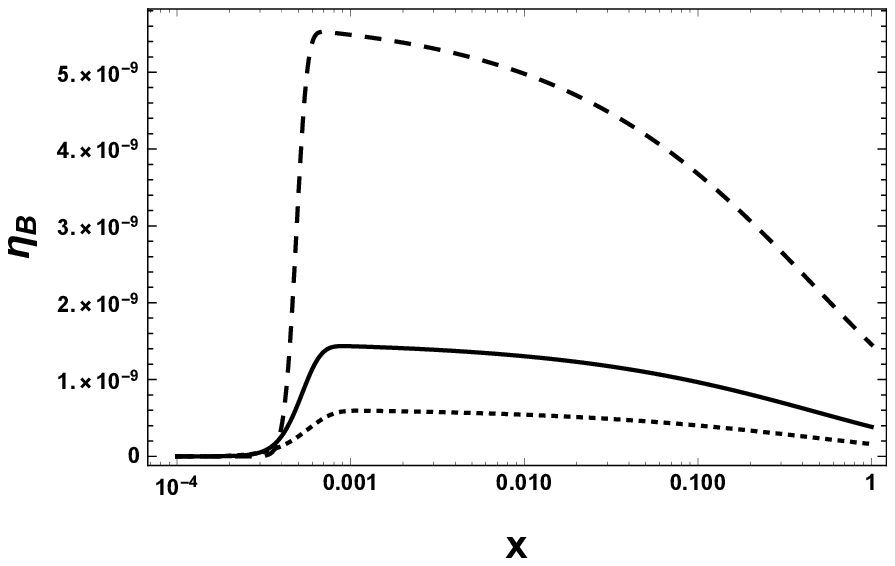}}
	\hspace{8mm}
	\subfigure[]{\label{fig:figure:211} 
		\includegraphics[width=.45\textwidth]{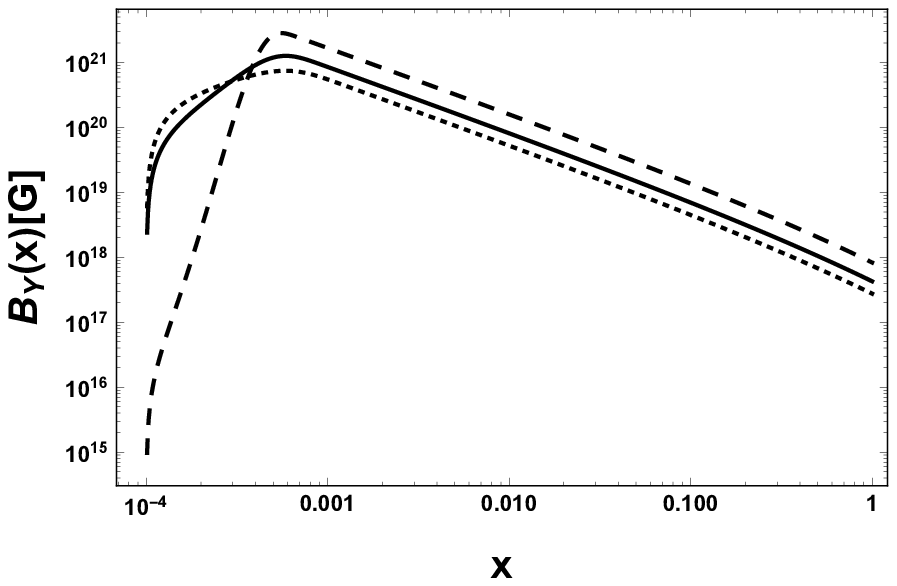}}
	\hspace{8mm}
	\caption{\footnotesize Time plots of: (a) the baryon asymmetry $\eta_{B}$, and (b) the hypermagnetic field amplitude $B_{Y}$, for various values of the width of fluctuations. The initial conditions are: $k=10^{-7}$, $B_{Y}^{(0)}=0$,  $\eta_{e_R}^{(0)}=\eta_{e_L}^{(0)}=\eta_{B}^{(0)}=0$, $v_0=10^{-5}$, $\beta_{0}=5\times10^{-4}$, and $x_{0}=45\times10^{-5}$. The dotted line is obtained for $b=3\times10^{-4}$, the solid line for $b=2\times10^{-4}$, the dashed line for $b=10^{-4}$.}
	\label{fig44}
\end{figure*}

For our second case we solve the set of evolution equations with the initial conditions, $v_0=10^{-5}$, $\beta_{0}=5\times10^{-4}$, and $x_{0}=45\times10^{-5}$, for various values of the width or duration of both fluctuations $b$, and show the results in Fig.\ \ref{fig44}. As can be seen, by decreasing the width of the Gaussian function, the maximum and the final values of the hypermagnetic field amplitude, and the baryon asymmetry increase.\footnote{We have also used a few other profiles, the results of which we can summarize as follows. For smooth profiles with the same normalization, the results are mainly dependent on the widths and not on the precise functional form of the profiles. However, the results usually increase by an order of magnitude for profiles functions which have discontinuities.}

\begin{figure*}[!ht]
	\subfigure[]{\label{fig:figure:im3} 
		\includegraphics[width=.45\textwidth]{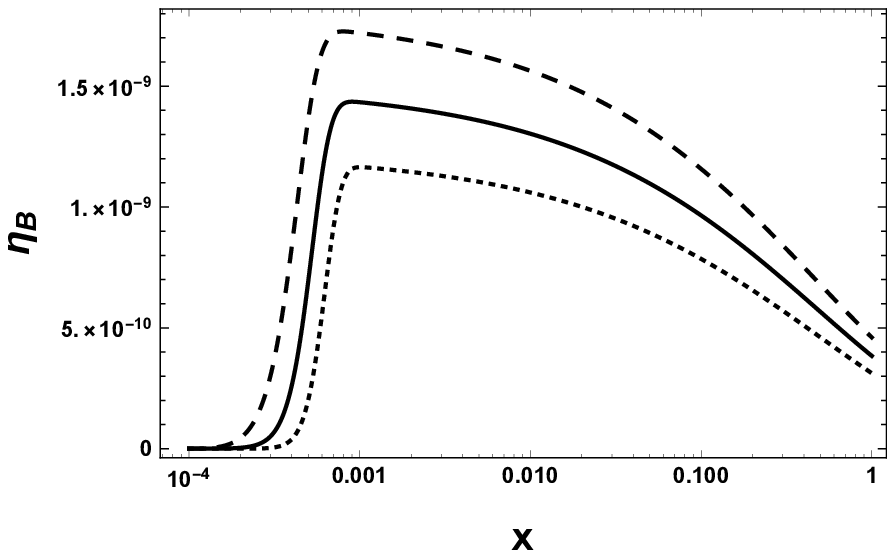}}
	\hspace{8mm}
	\subfigure[]{\label{fig:figure:in1}
		\includegraphics[width=.45\textwidth]{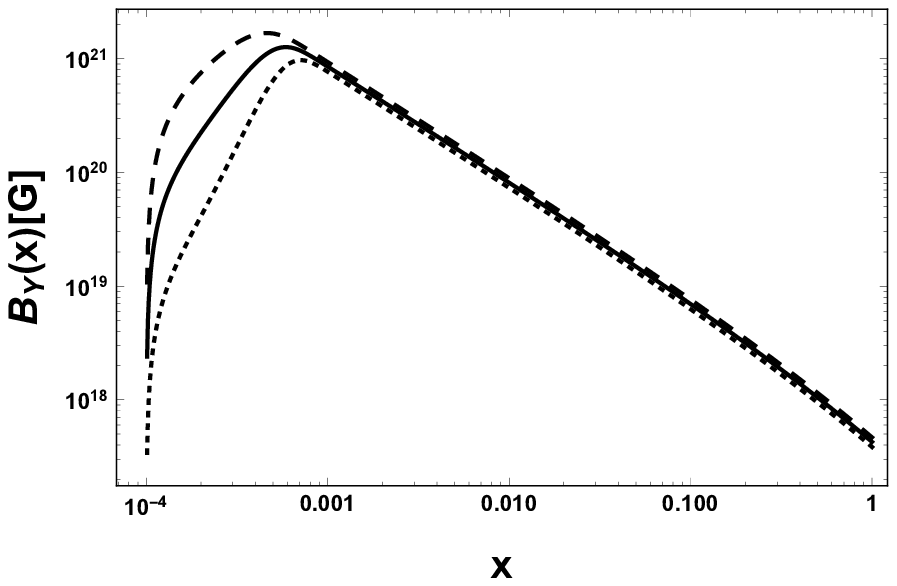}}
	\hspace{8mm}	
	\caption{\footnotesize Time plots of: (a) the baryon asymmetry $\eta_{B}$, and (b) the hypermagnetic field amplitude $B_{Y}$, for various values of the time of fluctuations. The initial conditions are: $k=10^{-7}$, $B_{Y}^{(0)}=0$, $\eta_{e_R}^{(0)}=\eta_{e_L}^{(0)}=\eta_{B}^{(0)}=0$, $v_0=10^{-5}$, $\beta_{0}=5\times10^{-4}$, $b=2\times10^{-4}$. The dotted line is obtained for $x_{0}=55\times10^{-5}$, the solid line for $x_{0}=45\times10^{-5}$, and the dashed line for $x_{0}=35\times10^{-5}$.}
	\label{fig4}
\end{figure*}

For our third case, we solve the coupled equations with the initial conditions, $v_0=10^{-5}$, $b=2\times10^{-4}$, and $\beta_{0}=5\times10^{-4}$, for various values of center time of the fluctuations $x_{0}$, and present the results in Fig.\ \ref{fig4}. As can be seen, when the fluctuations occur at an earlier time or higher temperature, the maxima and the final amplitudes of the hypermagnetic fields increase, and as a result, the matter-antimatter asymmetries increase as well.

\begin{figure*}[!ht]
	\subfigure[]{\label{fig:figure:1111}
		\includegraphics[width=.45\textwidth]{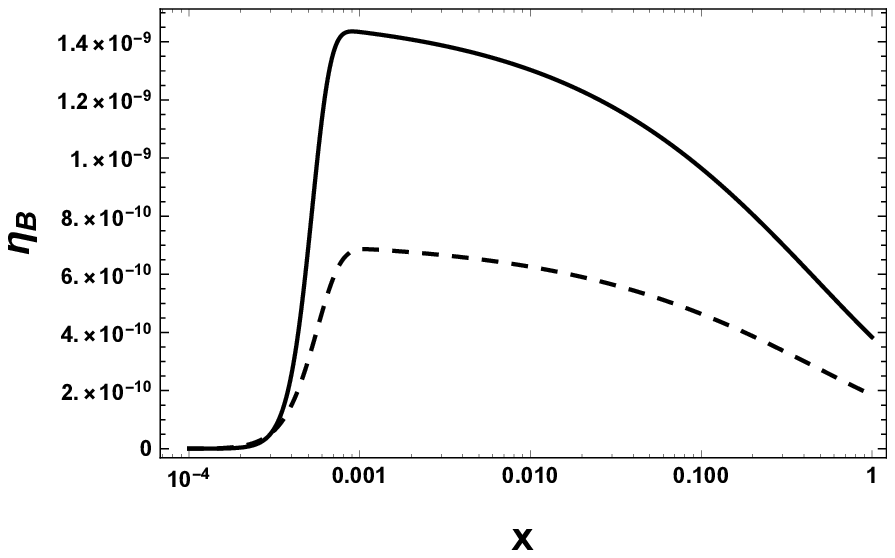}}
	\hspace{8mm}
	\subfigure[]{\label{fig:figure:2111} 
		\includegraphics[width=.45\textwidth]{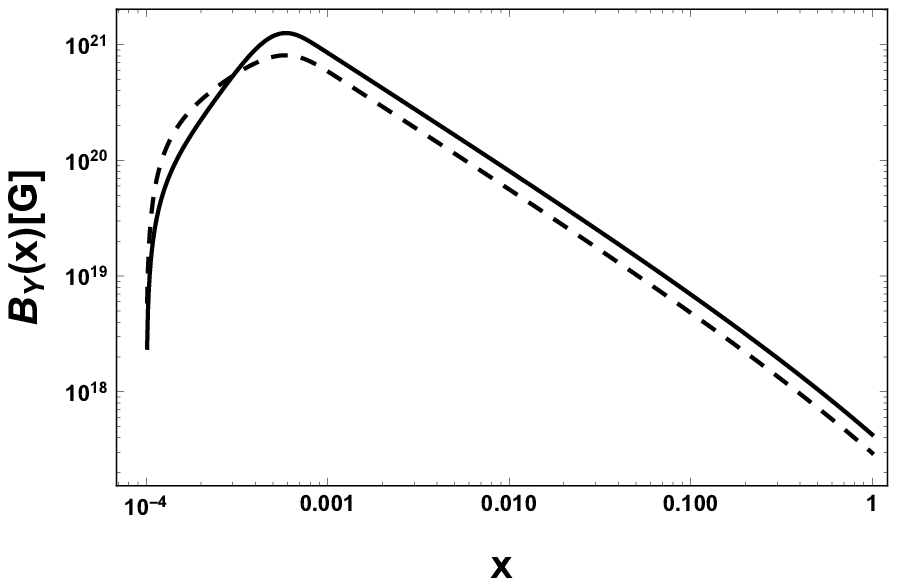}}
	\hspace{8mm}
	\caption{\footnotesize Time plots of: (a) the baryon asymmetry $\eta_{B}$, and (b) the hypermagnetic field amplitude $B_{Y}$, for two different vorticity configurations. The initial conditions are: $k=10^{-7}$, $B_{Y}^{(0)}=0$, $\eta_{e_R}^{(0)}=\eta_{e_L}^{(0)}=\eta_{B}^{(0)}=0$, $\beta_{0}=5\times10^{-4}$, $b=2\times10^{-4}$, and $x_{0}=45\times10^{-5}$. The solid line is for vorticity fluctuation with $v_0=10^{-5}$, and the dashed line is for constant vorticity with $v_0=10^{-2}$.}
	\label{fig441}
\end{figure*}

For our fourth and final case, we solve the set of evolution equations, with the initial conditions $\beta_{0}=5\times10^{-4}$, $b=2\times10^{-4}$ and $x_{0}=45\times10^{-5}$, for two different vorticity configurations. First configuration is a vorticity fluctuation with amplitude $v_0=10^{-5}$ and $x_{0}=45\times10^{-5}$. Second configuration is a constant vorticity with amplitude $v_0=10^{-2}$. The results are presented in Fig.\ \ref{fig441}. As can be seen from the figure, the general trends of the evolution curves are similar. The prominent feature of this comparison is the surprising result that a fluctuation with amplitude smaller by three orders of magnitude produces results comparable with the constant vorticity configuration.

Now we can address our assertion that $\mu/T\ll1$ within our model, i.e., our initial conditions and results. First, note that upon using the relations $\mu_{f}=(6s/T^2)\eta_{f}$ and $s=(2\pi^{2}g^{\star}/45)T^{3}$, we obtain $(\mu_{f}/T)=(12\pi^{2}g^{\star}/45)\eta_{f}=280.95 \eta_{f}$. The largest asymmetry in our results is obtained for $\eta_{B}$ and is shown in the Fig.\ \ref{fig:figure:111}. This figure shows that $(\eta_{B})_{\mathrm{max}}\sim 5\times10^{-9}$, leading to $(\mu/T)_{\mathrm{max}}\sim 10^{-6}$. The condition $\mu/T\ll1$ has been used, for example, to justify neglecting terms of $O(\mu/T)$ in Eqs.~(\ref{a-3}-\ref{a-6}) in Appendix A, leading to the expressions for $c_{\mathrm{v}}$ and $c_{\mathrm{B}}$ shown in Eqs.~(\ref{eq22},\ref{eqc_B}), respectively. Moreover, as shown in Appendix B, this condition, along with the assumption of non-relativistic velocity of the plasma, imply that the CME and CVE contributions to the temporal components of the four-currents in the AMHD equations are negligible within our model.

\begin{figure}[H]
		\centering
		\subfigure[]{\label{fig:figure:1.1.311sa}
			\includegraphics[width=.45\textwidth]{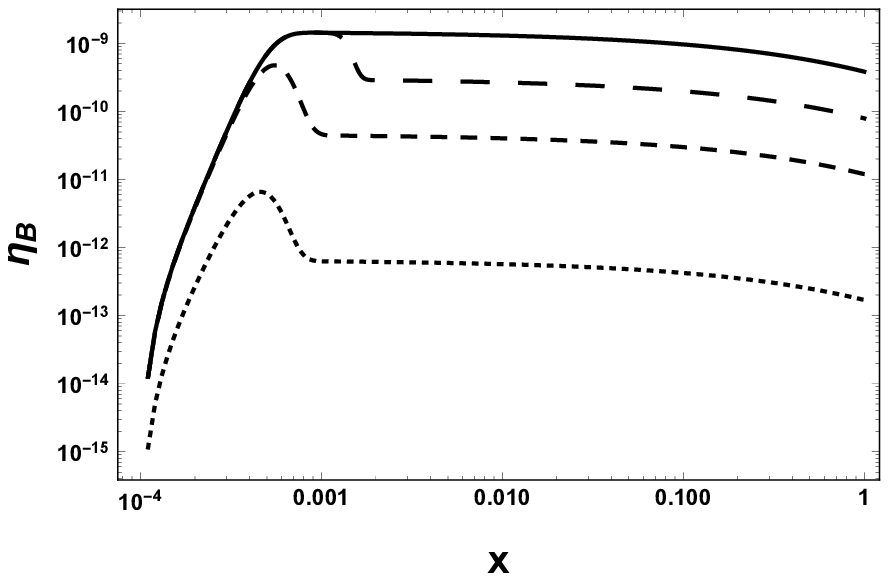}}
		\hspace{8mm}
		\subfigure[]{\label{fig:figure:1.1.211sa}
			\includegraphics[width=.45\textwidth]{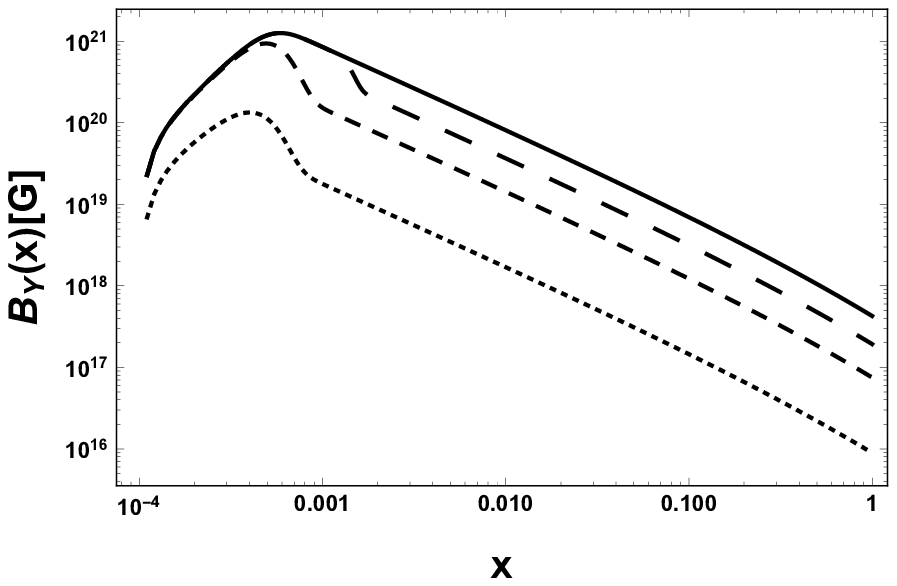}}
		\hspace{8mm}
\caption{\footnotesize 	Time plots of: (a)  the baryon asymmetry $\eta_{B}$, (b)  the hypermagnetic field amplitude $B_{Y}$ for two sets of successive and opposing fluctuations. The initial conditions are:  $k=10^{-7}$, $B_{Y}^{(0)}=0$,  $\eta_{e_R}^{(0)}=\eta_{e_L}^{(0)}=\eta_{B}^{(0)}=0$, $v_{0,+}=v_{0,-}=10^{-5}$, $b=2\times10^{-4}$, $\beta_{0,+}=-\beta_{0,-}=5\times10^{-4}$, and  $x_{0,+}=4.5\times10^{-4}$. The large dashed  line is for $x_{0,-}=1.45\times10^{-3}=5b+x_{0,+}$,  the medium dashed line is for $x_{0,-}=6.5\times10^{-4}=b+x_{0,+}$, the dotted line is for $x_{0,-}=4.7\times10^{-4}=0.1b+x_{0,+}$, and the solid line is obtained in the absence of the second set of fluctuations.}\label{fig1.11sa}
\end{figure}

We finally address the issue of two successive pulses with opposite temperature profiles to see by how much can the second pulse negate the results of the first. For this purpose it suffices to assume that the profile of the vorticities are unchanged. To be specific, we assume
	\begin{equation}\label{Tprofile22}
		\beta(x)=\beta_{+}(x)+\beta_{-}(x), 
	\end{equation}
	\begin{equation}\label{Tprofile222}
		v(x)=v_{+}(x)+v_{-}(x), 
	\end{equation}
	where
	\begin{equation}\label{Tprofile2222}
		\beta_{\pm}(x)=\frac{\pm\beta_{0}}{b\sqrt{2\pi}}\exp\left[-\frac{(x-x_{0,\pm})^{2}}{2b^{2}}\right], 
	\end{equation}
	and
	\begin{equation}\label{Tprofile1}
		v_{\pm}(x)=\frac{v_{0}}{b\sqrt{2\pi}}\exp\left[-\frac{(x-x_{0,\pm})^{2}}{2b^{2}}\right]. 
	\end{equation}
We consider three cases in which the time separation of the pulses $\Delta x_0= x_{0,+}-x_{0,-}$ are $5b$, $b$ and $0.1b$, where $b$ denotes the width of the pulses. The results are shown in Fig.\ \ref{fig1.11sa}, where we also show our results for a single pulse for comparison. As can be seen, the final values of the asymmetries generated are reduced, as compared to the single pulse case, by a factor of about 5, 50 and 1000, respectively. The final values of the hypermagnetic field generated are reduced by square root of values stated above. It is interesting to note that even in the case $\Delta x_0=0.1b$, the model produced $\eta_B \simeq 10^{-13}$, a value which can be increased easily by increasing $\beta_0$, $v_0$, and/or decreasing $b$.


\section{Conclusion}\label{x5}
\noindent In this study, we have investigated the contribution of the temperature-dependent CVE to the generation and evolution of the hypermagnetic fields and the matter-antimatter asymmetries, in the symmetric phase of the early Universe and in the temperature range $100\mbox{ GeV} \le T\le 10\mbox{ TeV}$. The CVE has two possible sources in a vortical plasma, one from the chiralities and the other from the temperature of the particles. The former has been investigated in the literature much more than the latter. Here, we have focused on the latter in the form of transient temperature fluctuations, and have shown its important role in the production and evolution of the hypermagnetic fields and the matter-antimatter asymmetries. The transient fluctuations that we have considered are in the form of sharp Gaussian shaped pulses. In particular, we have shown that small simultaneous and transient fluctuations of vorticity about zero background value and temperature of some matter degrees of freedom about the equilibrium temperature of the plasma, close to the EWPT, can generate strong hypermagnetic fields and large  matter-antimatter asymmetries, even in the absence of any initial seed for the hypermagnetic field or any initial matter-antimatter asymmetries. Furthermore, we have shown that, an increase in the amplitude of temperature or vorticity fluctuations leads to the production of stronger hypermagnetic fields, and therefore, larger matter-antimatter asymmetries. This outcome has not been observed in any of the previous studies. In some studies which only take the CME into account, either an initially strong hypermagnetic field produces matter-antimatter asymmetries, or initial large matter-antimatter asymmetries strengthen a preexisting seed of hypermagnetic field \cite{sh1,sh2,sh3}. In some other studies which also include the CVE and assume large initial chiralities in the vortical plasma, a seed of the hypermagnetic field is produced which then grows due to the CME \cite{Abbaslu1}.

In this work, we have considered a simple monochromatic helical configuration for the vorticity and hypermagnetic fields with a negative helicity, which ensures the production of the desired positive matter-antimatter asymmetries. Furthermore, we have shown that, either an increase in the amplitude of the temperature or vorticity fluctuations, or a decrease in their widths leads to the production of stronger hypermagnetic fields, and therefore, larger matter-antimatter asymmetries. We have also shown that fluctuations in vorticity are several orders of magnitude more productive than constant vorticity. Within our model, the temperature-dependent CVE is the dominant effect as compared to the CME. 
We have also shown that when there are two sets of successive pulses with opposite temperature profiles,  the first pulse is the main determinant of the outcome. A generalization of this work would be a stochastic analysis of fluctuations of both temperature and vorticity.

\section{APPENDIX A}

In this appendix we present the expressions for the chiral vorticity and helicity coefficients, $c_{\mathrm{v}}$ and $c_{\mathrm{B}}$ given by Eqs.\ (\ref{eq22}), and (\ref{eqc_B}). First, we start with the relevant and well known expressions in the broken phase. In relativistic hydrodynamics, the energy and number currents can flow separately in the presence of dissipative processes, therefore the definition of the flow will not be trivial\cite{2019jkc}. 
Some common hydrodynamic frames in which the equations of AMHD may be formulated include the Landau-Lifshitz (or energy) frame \cite{Landau-Lif}, the Eckart (or conserved charge/particle) frame \cite{Eckart-1940zz}, and the more recently introduced anomalous ``no-drag'' frame\cite{Stephanov-2015r}. In the latter, as its name suggests, a stationary obstacle experiences no drag, even  when the energy and charge currents are present. In the Landau-Lifshitz frame, the energy-momentum tensor $T^{\mu\nu}$ and the total electric current $J^{\mu}$ for a plasma consisting of a single species of massless fermions  of both chiralities are given by 
\begin{equation}
T^{\mu\nu}=(\rho+p)u^{\mu}u^{\nu}- p g^{\mu\nu}+\frac{1}{4}g^{\mu\nu} F^{\alpha\beta} F_{\alpha\beta}-F^{\nu\sigma}{F^{\mu}}_{\sigma}+ \tau^{\mu\nu},
\end{equation}

\begin{equation}
J^{\mu}=\rho_{\mathrm{el}} u^{\mu}+J^{\mu}_\mathrm{cm}+J^{\mu}_\mathrm{cv}+\nu^{\mu},
\end{equation}
\begin{equation}
J^{\mu}_\mathrm{cm}=(Q_\mathrm{R}\xi_{\mathrm {B,R}}+Q_\mathrm{L}\xi_{\mathrm {B,L}})B^{\mu}=c_{\mathrm{B}}B^{\mu},
\end{equation}
\begin{equation}
J^{\mu}_\mathrm{cv}=(Q_\mathrm{R}\xi_{\mathrm {v,R}}+Q_\mathrm{L}\xi_{\mathrm {v,L}})\omega^{\mu}=c_{\mathrm{v}}\omega^{\mu},
\end{equation}
where $F_{\alpha\beta}=\nabla_{\alpha}A_{\beta}-\nabla_{\beta}A_{\alpha}$ is the field strength tensor, $p$ and $\rho$ are the pressure and the energy density of the plasma, $\rho_{\mathrm{el}}$ is the electric charge density, $u^{\mu}=\gamma\left(1,\vec{v}/R\right)$ is the four-velocity of the plasma normalized such that $u^{\mu}u_{\mu}=1$, and $\gamma$ is the Lorentz factor. \footnote{Note that the self-consistency of our calculation in which the diagonal Einstein tensor obtained from the FRW metric is used implies that not only should the electromagnetic field density be small compared to the energy density of the Universe \cite{b1234ss}, but also the bulk velocity should obey the condition $\left |\vec{v}\right | \ll1$, or equivalently $\gamma\simeq1$ and $u^{\mu}\simeq(1,\vec{v}/R)$. } In the above equations, $\nu^{\mu}$ and $\tau^{\mu\nu}$ denote the electric diffusion current and viscous stress tensor, respectively \cite{son1}, $Q_{R}$ ($Q_{L}$)  denotes the electric charges of the right-handed (left-handed) fermions,  $B^{\mu}=(\epsilon^{\mu\nu\rho\sigma}/2R^3)u_{\nu}F_{\rho\sigma}$ is the magnetic field four-vector, and $\omega^{\mu}=(\epsilon^{\mu\nu\rho\sigma}/R^3)u_{\nu}\nabla_{\rho}u_{\sigma}$ is the vorticity four-vector, with the totally
anti-symmetric Levi-Civita tensor density specified by $\epsilon^{0123}=-\epsilon_{0123}=1$.


Furthermore, in the  Landau-Lifshitz frame, the CME and CVE coefficients for chiral fermions are given as \cite{son1,Yamamoto16,Anand:2017,Landsteiner-2016,Neiman-2010zi,Landsteiner-2012kdw}

\begin{equation}\label{a-3}
\xi_{\mathrm {B,R}}=\frac{Q_\mathrm{R}\mu_\mathrm{R}}{4\pi^2}\Big[1-\frac{1}{2}\frac{n_\mathrm{R}\mu_\mathrm{R}}{\rho+p}\Big]-\frac{1}{24}\frac{n_\mathrm{R}T^{2}}{\rho+p},
\end{equation}

\begin{equation}\label{a-4}
\xi_\mathrm{B,L}=
-\frac{Q_\mathrm{L}\mu_\mathrm{L}}{4\pi^2}\Big[1-\frac{1}{2}\frac{n_\mathrm{L}\mu_\mathrm{L}}{\rho+p}\Big]+\frac{1}{24}\frac{n_\mathrm{L}T^{2}}{\rho+p},
\end{equation}

\begin{equation}\label{a-5}
\xi_{\mathrm {v,R}}=\frac{\mu_\mathrm{R}^{2}}{8\pi^2}\Big[1-\frac{2}{3}\frac{n_\mathrm{R}\mu_\mathrm{R}}{\rho+p}\Big]+\frac{1}{24}T^{2}\Big[1-\frac{2n_\mathrm{R}\mu_\mathrm{R}}{\rho+p}\Big],
\end{equation}

\begin{equation}\label{a-6}
\xi_{\mathrm {v,L}}=-\frac{\mu_\mathrm{L}^{2}}{8\pi^2}\Big[1-\frac{2}{3}\frac{n_\mathrm{L}\mu_\mathrm{L}}{\rho+p}\Big]-\frac{1}{24}T^{2}\Big[1-\frac{2n_\mathrm{L}\mu_\mathrm{L}}{\rho+p}\Big].
\end{equation}
where $T$ is the temperature and  $n_\mathrm{R}$ ($n_\mathrm{L}$) is the right-handed (left-handed) charge density .\footnote{The charge density $n$ is the difference between the particle and anti-particle charge densities.} As we shall show explicitly in Sec.\ \ref{x4}, $\mu_\mathrm{R,L}/T\ll1$ within our model, i.e., our initial conditions and results.  Hence, $n_\mathrm{R,L}\simeq\frac{1}{6}\mu_\mathrm{R,L}T^{2}$ and $\rho=3p\simeq \frac{\pi^{2}}{30}g^{*}T^4$,  and Eqs.\ (\ref{a-3}-\ref{a-6}) may be simplified as follows 

\begin{equation}\label{a-32}
\xi_{\mathrm {B,R}}\simeq\frac{Q_\mathrm{R}\mu_\mathrm{R}}{4\pi^2},
\end{equation}

\begin{equation}\label{a-42}
\xi_\mathrm{B,L}\simeq
-\frac{Q_\mathrm{L}\mu_\mathrm{L}}{4\pi^2},
\end{equation}

\begin{equation}\label{a-52}
\xi_{\mathrm {v,R}}\simeq\frac{\mu_\mathrm{R}^{2}}{8\pi^2}+\frac{1}{24}T^{2},
\end{equation}

\begin{equation}\label{a-62}
\xi_{\mathrm {v,L}}\simeq-\frac{\mu_\mathrm{L}^{2}}{8\pi^2}-\frac{1}{24}T^{2}.
\end{equation}
Moreover, since $\mu/T\ll1$, we will consider only the Ohmic part of the diffusion current $\nu^{\mu}=\sigma [E^{\mu}+T(u^{\mu}u^{\nu}-g^{\mu\nu})\nabla_{\nu}\big(\frac{\mu}{T}\big)]$ given by the first term, where $\sigma=\sigma_{R}+\sigma_{L}$ is the electrical conductivity, and $E^{\mu}=F^{\mu\nu}u_{\nu}$ is the electric field four-vector.
In order to carry over these results to the symmetric phase of the early Universe, it suffices to replace the electromagnetic field by the hypercharge gauge field and the electric charges of different particle species by their relevant hypercharges. Taking into account all three generations of leptons and quarks, we can easily obtain the chiral vorticity and helicity coefficients $c_{\mathrm{v}}$ and $c_{\mathrm{B}}$, given by Eqs.\ (\ref{eq22}), and (\ref{eqc_B}), using Eqs.\ (\ref{a-32}-\ref{a-62}).  The four-vectors $B^{\mu}$, $\omega^{\mu}$ , $a^{\mu}$, and $E^{\mu}$ in the limit $v\ll1$ are given below,
\begin{equation}
	B^{\mu}=\gamma\left(\vec{v}.\vec{B},\frac{\vec{B}-\vec{v}\times\vec{E}}{R}\right)\simeq\left(\vec{v}. \vec{B},\frac{\vec{B}}{R}\right)
\end{equation}
\begin{equation}
		\omega^{\mu}=\gamma\left(\vec{v}.\vec{\omega},\frac{\vec{\omega}-\vec{v}\times\vec{a}}{R}\right)
		\simeq	\left(\vec{v}.\vec{\omega},\frac{\vec{\omega}}{R}\right)
	\end{equation}
	\begin{equation}
 a^\mu=\gamma\left(\vec{v}.\vec{a},\frac{\vec{a}+\vec{v}\times\vec{\omega}}{R}\right) \simeq \left(\vec{v}.\vec{a},\frac{\vec{a}}{R}\right)
	\end{equation}
\begin{equation}
	E^{\mu}=\gamma\left(\vec{v}.\vec{E},\frac{\vec{E}+\vec{v}\times\vec{B}}{R}\right) \simeq \left(\vec{v}.\vec{E},\frac{\vec{E}+\vec{v}\times\vec{B}}{R}\right),
\end{equation}
where $a^\mu=\Omega^{\mu\nu}u_\nu$ is acceleration four-vector, $\Omega_{\mu\nu}=\nabla_{\mu}u_\nu-\nabla_{\nu}u_\mu$ is vorticity tensor, $a^{i}=R \Omega^{0i}$ is three vector acceleration. We have also used the assumption $\partial_{t}\sim\vec{\nabla}.\vec{v}$ in the derivative expansion of the hydrodynamics, so  $\vec{v}\times\vec{E}\simeq v^{2}\vec{B}$ and we have ignored the terms of $O(v^2)$. 


\section{APPENDIX B}

In this appendix we start with the basic expression for the anomaly given by Eq.\ (\ref{anomaly1}) and obtain its explicit form which eventually leads to Eqs.\ (\ref{anomaly2},\ref{eq47},\ref{eq48}). The equations of AMHD consist of energy-momentum conservation, Maxwell's equations and the anomaly relations. These equations may be expressed covariantly as
\begin{equation}
\nabla_{\mu}T^{\mu\nu}= 0,
\end{equation}
\begin{equation}
\begin{split}
&\nabla_{\mu}F^{\mu\nu}=J^{\nu}\\&
\nabla_{\mu}{\tilde{F}}^{\mu\nu}=0
\end{split}
\end{equation}
\begin{equation}\label{eq-ab1}
\nabla_{\mu}j^{\mu}_\mathrm{R,L}=C_\mathrm{R,L} E_{\mu}B^{\mu},
\end{equation}
where ${\tilde{F}}^{\mu\nu}=\frac{1}{2R^3}\epsilon^{\mu\nu\rho\sigma}F_{\rho\sigma}$ is the dual field strength tensor, $C_{R,L}$ are the corresponding right- and left- handed anomaly coefficients, and $j^{\mu}_{R,L}$ are the fermionic currents given as follows

\begin{equation}\label{eq-ab2}
\begin{split}
&j^{\mu}_\mathrm{R}=n_\mathrm{R} u^{\mu}+\xi_{\mathrm {B,R}}B^{\mu}+\xi_{\mathrm {v,R}}\omega^{\mu}+V^{\mu}_{R},\\&
j^{\mu}_\mathrm{L}=n_\mathrm{L} u^{\mu}+\xi_{\mathrm {B,L}}B^{\mu}+\xi_{\mathrm {v,L}}\omega^{\mu}+V^{\mu}_{L},
\end{split} 
\end{equation} 
where $n_\mathrm{R,L}$  and $V^{\mu}_\mathrm{R,L}$ are the chiral charge density  and chiral particle diffusion current, respectively \cite{son1,Yamamoto16,Anand:2017,Landsteiner-2016}. The anomaly equations given above can be written out as

\begin{equation}\label{eq4aq2}
\partial_{t}j^{0}_\mathrm{(R,L)}+ \frac{1}{R}\vec{\nabla}.\vec{j}_{(R,L)}+ 3Hj^{0}_\mathrm{(R,L)} =C_\mathrm{(R,L)} E_{\mu}B^{\mu},
\end{equation} 
where 
 \begin{equation}\label{tcomp}
j^{0}_{(R,L)}=n_\mathrm{(R,L)}+\xi_{\mathrm {B,(R,L)}}\vec{v}.\vec{B}+\xi_{\mathrm {v,(R,L)}}\vec{v}.\vec{\omega}+V^{0}_{R,L}.
\end{equation} 

	Upon taking the spatial average of Eq.\ (\ref{eq4aq2}), the boundary term vanishes and we obtain\footnote{Note that $E_{\mu}B^{\mu}=(\vec{v}.\vec{E})(\vec{v}.\vec{B})-(\vec{E}+\vec{v}\times\vec{B}).(\vec{B}-\vec{v}\times\vec{E})= -\vec{E}.\vec{B}+O(v^2)$, and we have ignored the terms of $O(v^2)$.}
\begin{equation}\label{eq4aq3w}
\begin{split}
\partial_{t}n_\mathrm{(R,L)}+  3H n_\mathrm{(R,L)} =&-\Big[\partial_{t}+3H\Big]\Big[\xi_\mathrm{B,(R,L)}\langle\vec{v}.\vec{B}\rangle+\xi_\mathrm{v,(R,L)}\langle\vec{v}.\vec{\omega}\rangle+\frac{\sigma_\mathrm{R,L}}{Q_\mathrm{R,L}}\langle\vec{v}.\vec{E}\rangle\Big] \\&-C_{R,L} \langle\vec{E}.\vec{B}\rangle.
\end{split}
\end{equation}
Using the relation $\dot{s}/s=-3H$, the anomaly equation reduces to the simplified form
\begin{equation}\label{eq4aq3w1}
\begin{split}
 \partial_{t}\left(\frac{n_\mathrm{(R,L)}}{s}\right)= &-\partial_{t}\Big[\frac{\xi_\mathrm{B,(R,L)}}{s}\langle\vec{v}.\vec{B}\rangle+\frac{\xi_\mathrm{v,(R,L)}}{s}\langle\vec{v}.\vec{\omega}\rangle+\frac{\sigma_\mathrm{R,L}}{Q_\mathrm{R,L}s}\langle\vec{v}.\vec{E}\rangle\Big]-\frac{C_\mathrm{R,L}}{s} \langle\vec{E}.\vec{B}\rangle.
\end{split}
\end{equation}
It is worth mentioning that since $(\partial_{t}+3H)\langle\vec{A}.\vec{B}\rangle=-2\langle\vec{E}.\vec{B}\rangle$, Eq.\ (\ref{eq4aq3w1}) can be rewitten in the following form
\begin{equation}\label{eq4aq3wsaw1efs}
\begin{split}
\partial_{t}\Big[\eta_{R,L}+ \frac{\xi_{\mathrm {B,(R,L)}}}{s}\langle\vec{v}.\vec{B}\rangle+\frac{\xi_{\mathrm {v,(R,L)}}}{s}\langle\vec{v}.\vec{\omega}\rangle+\frac{\sigma_{(R,L)}}{Q_{(R,L)}s}\langle\vec{v}.\vec{E}\rangle-\frac{C_\mathrm{R,L}}{2s}\langle\vec{A}.\vec{B}\rangle\Big]=0,
\end{split}
\end{equation} 
which clearly reveals the conservation of a generalized charge \cite{a3,Yamamoto16}.

Using Eqs.\ (\ref{eq4aq3w1}, \ref{a-32}-\ref{a-62}) and $x=(T_\mathrm{EW}/T)^2=t/t_\mathrm{EW}$, we obtain 
\begin{equation}
\begin{split}
\frac{d\eta_\mathrm{R}}{dx}=&\frac{1}{\Big[1+\frac{6Q_\mathrm{R}}{4\pi^{2}}\frac{\langle\vec{v}.\vec{B}\rangle}{10^{20}G}\frac{x}{5000}-\frac{36M}{4\pi^{2}}\eta_\mathrm{R}kv^{2}\Big]}\Bigg[-\frac{6Q_\mathrm{R}}{4\pi^{2}}\frac{x}{5000}\frac{\eta_\mathrm{R}}{10^{20 }G}\Big[\langle\vec{v}.\partial_{x}\vec{B}\rangle+\langle\frac{\vec{v}.\vec{B}}{x}\rangle+\langle\vec{B}.\partial_{x}\vec{v}\rangle\Big]\\&+\Big[\frac{36M}{4\pi^{2}}\eta_\mathrm{R}^{2}+\frac{1}{12M}\Big]k\vec{v}.\partial_{x}\vec{v}-\frac{x}{50MQ_\mathrm{R}10^{20}  G}\Big[\langle\vec{v}.\partial_{x}\vec{E}\rangle+\langle\frac{\vec{v}.\vec{E}}{x}\rangle+\langle\vec{E}.\partial_{x}\vec{v}\rangle\Big]\\&-\frac{t_\mathrm{EW}C_\mathrm{R}}{s}\langle\vec{E}.\vec{B}\rangle\Bigg],
\end{split}
\end{equation} 
\begin{equation}
\begin{split}
	\frac{d\eta_\mathrm{L}}{dx}=&\frac{1}{\Big[1-\frac{6Q_\mathrm{L}}{4\pi^{2}}\frac{\langle\vec{v}.\vec{B}\rangle}{10^{20}G}\frac{x}{5000}+\frac{36M}{4\pi^{2}}\eta_\mathrm{L}kv^{2}\Big]}\Bigg[\frac{6Q_\mathrm{L}}{4\pi^{2}}\frac{x}{5000}\frac{\eta_\mathrm{L}}{10^{20 }G}\Big[\langle\vec{v}.\partial_{x}\vec{B}\rangle+\langle\frac{\vec{v}.\vec{B}}{x}\rangle+\langle\vec{B}.\partial_{x}\vec{v}\rangle\Big]\\&-\Big[\frac{36M}{4\pi^{2}}\eta_\mathrm{L}^{2}+\frac{1}{12M}\Big]k\vec{v}.\partial_{x}\vec{v}-\frac{x}{50MQ_\mathrm{L}10^{20}  G}\Big[\langle\vec{v}.\partial_{x}\vec{E}\rangle+\langle\frac{\vec{v}.\vec{E}}{x}\rangle+\langle\vec{E}.\partial_{x}\vec{v}\rangle\Big]\\&-\frac{t_\mathrm{EW}C_\mathrm{L}}{s}\langle\vec{E}.\vec{B}\rangle\Bigg],
\end{split}
\end{equation} 
where $M=2\pi^{2}g^{*}/45$. Here the electric field $\vec{E}$ is not an independent variable and is determined by Eq.\ (\ref{eq16}). From the approximation of a vanishing displacement current, we have $\partial_{x}\vec{E}=-\vec{ E}/x$. 

\begin{figure*}[!ht]
		\subfigure[]{\label{fig:figure:1j}
			\includegraphics[width=.45\textwidth]{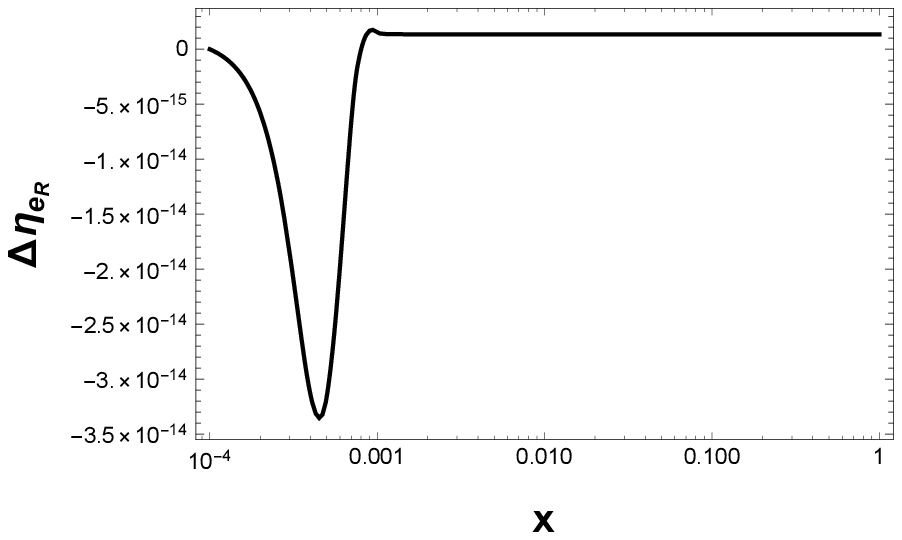}}
		\hspace{8mm}
		\subfigure[]{\label{fig:figure:2j}
			\includegraphics[width=.45\textwidth]{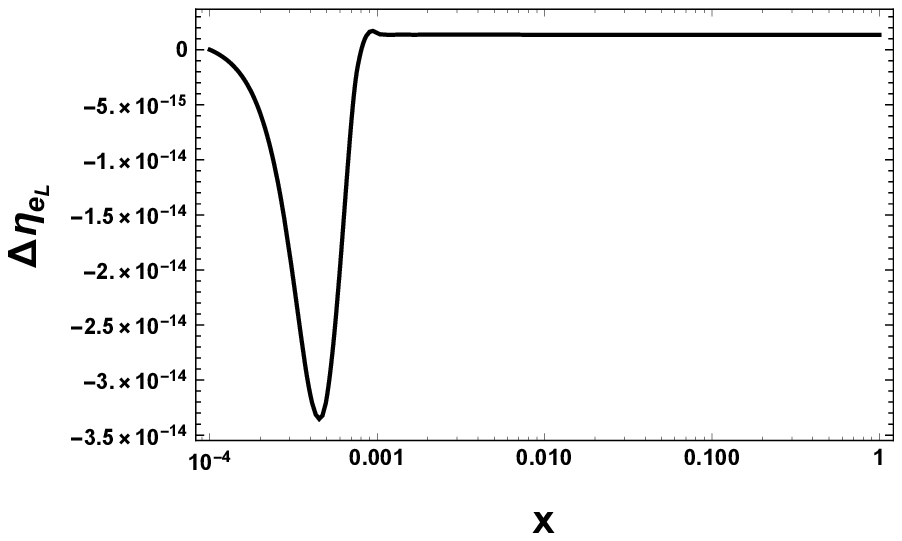}}
		\hspace{8mm}
		\subfigure[]{\label{fig:figure:11j}
			\includegraphics[width=.45\textwidth]{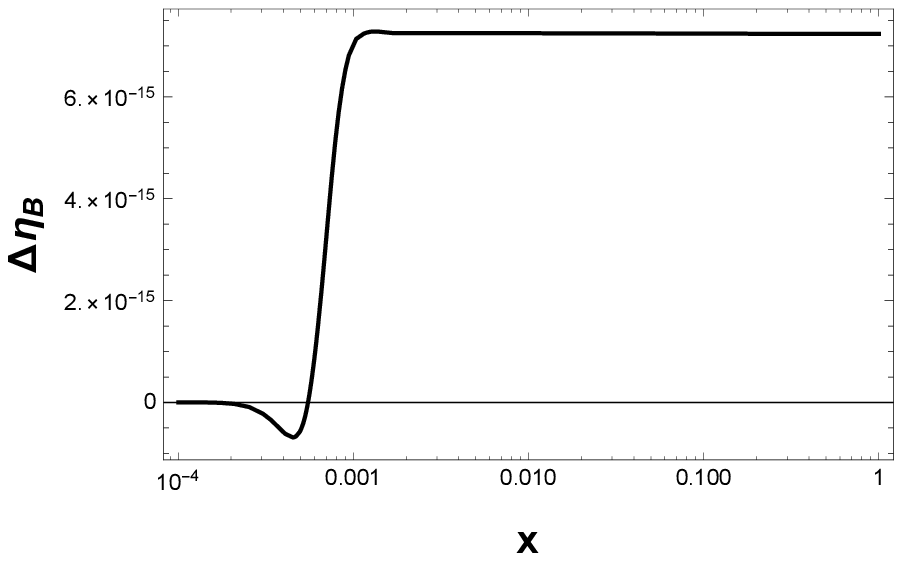}}
		\hspace{8mm}
		\subfigure[]{\label{fig:figure:21j} 
			\includegraphics[width=.45\textwidth]{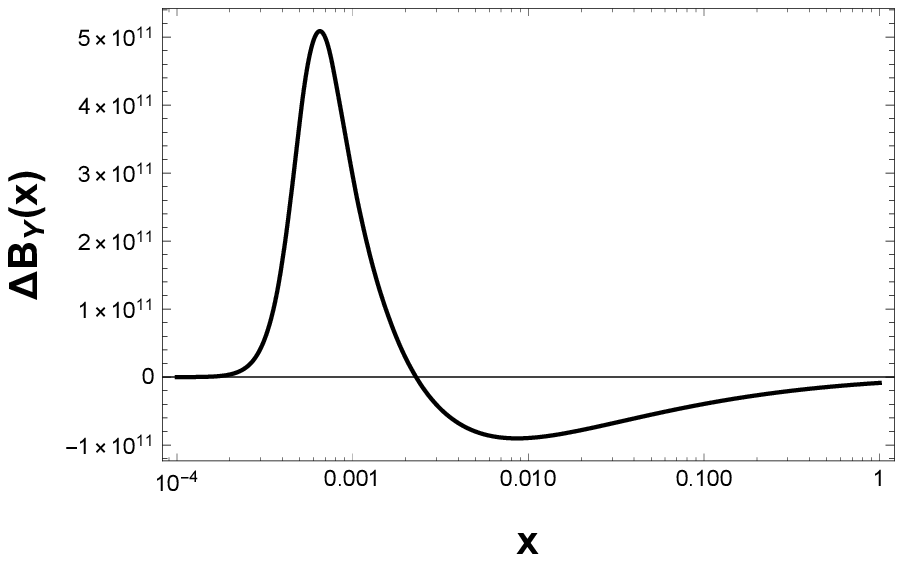}}
		\hspace{8mm}
		\caption{\footnotesize Time plots of the differences between the results with and without the inclusion of the temporal components of the CVE, CME and diffusion currents given by the last three terms of Eq.~(\ref{tcomp}): (a) the right-handed electron asymmetry $\Delta\eta_{e_{R}}$, (b) the left-handed electron asymmetry $\Delta\eta_{e_{L}}$, (c) the baryon asymmetry $\Delta\eta_{B}$, and (d) the hypermagnetic field amplitude $\Delta B_{Y}$. The initial conditions are: $k=10^{-7}$, $B_{Y}^{(0)}=0$, $\eta_{e_R}^{(0)}=\eta_{e_L}^{(0)}=\eta_{B}^{(0)}=0$, $\beta_{0}=5\times10^{-4}$, $v_0=10^{-5}$, $b=2\times10^{-4}$, and $x_{0}=45\times10^{-5}$. "}
		\label{fig4-1}
\end{figure*}

As is well known, in the symmetric phase of the early Universe $T>100$ GeV, and in Sec.\ \ref{x4} we show that the condition $\mu/T\leq 10^{-6}$ holds within our model. These conditions, in conjunction with the low-velocity limit considered in this work,
highly suppress the contributions of the temporal components of the CVE, CME and diffusion currents. To ascertain this claim, we have solved Eqs.\ (\ref{eq47})-(\ref{eq49}) both with and without the additional temporal components and present the differences in Fig.\ \ref{fig4-1}. Comparing the scales of this figure and Fig.\ref{fig1}, we observe the following for the differences: $\Delta\eta/\eta\sim 10^{-4}$  and $\Delta B_Y/B_Y\sim 10^{-9}$.

\end{document}